\documentclass[3p,12pt]{elsarticle}

\usepackage{amsmath}
\usepackage{amssymb}
\usepackage{color}
\usepackage{graphicx}
\usepackage{epsfig}
\usepackage{amsfonts}
\usepackage{float}
\usepackage{epstopdf}
\usepackage{mathrsfs}
\usepackage[%
breaklinks%
,colorlinks%
,linkcolor=red
,anchorcolor=blue%
,pagecolor=blue%
,citecolor=blue%
,bookmarks=ture%
]{hyperref}

\newtheorem{theorem}{Theorem}[section]

\newtheorem{lemma}[theorem]{Lemma}

\newtheorem{definition}[theorem]{Definition}
\newtheorem{remark}[theorem]{Remark}
\newtheorem{example}[theorem]{Example}
\newtheorem{alg}[theorem]{Algorithm}
\numberwithin{equation}{section}
\newproof{proof}{Proof}

\def\C{{\mathcal{C}}}

\def\N{{\mathbb{N}}}

\def\a{{\bar a}}
\def\b{{\bar b}}
\def\c{{\bar c}}
\def\d{{\bar d}}

\def\s{\bar{s}}
\def\u{\bar{u}}

\def\G{\mathscr{G}}

\def\PP{{\mathcal P}}
\def\PE{{\mathcal E}}

\def\S{{\mathbf{S}}}

\def\Q{{\mathbb{Q}}}

\begin{document}

\begin{frontmatter}
\title{Homeomorphic approximation of the intersection curve of two rational surfaces}

\author[SLY]{Liyong Shen}
 \ead{shenly@amss.ac.cn}
 \address[SLY]{School of Mathematical Sciences, Graduate University of
 CAS, Beijing, China}

 \author[JSC]{Jin-San Cheng}
 \ead{jcheng@amss.ac.cn}
\address[JSC]{Key Laboratory of Mathematics Mechanization, AMSS, CAS,
        Beijing, China}

 \author[JSC]{Xiaohong Jia}
 \ead{xhjia@amss.ac.cn}

\begin{abstract}
We present an approach of computing the intersection curve $\mathcal{C}$ of two rational parametric surface $\S_1(u,s)$ and $\S_2(v,t)$,  one being projectable and hence can easily be implicitized. Plugging the parametric surface to the implicit surface yields a plane algebraic curve $G(v,t)=0$. By analyzing the topology graph $\G$ of $G(v,t)=0$ and the singular points on the intersection curve $\mathcal{C}$ we associate a space topology graph to $\mathcal{C}$, which is homeomorphic to $\mathcal{C}$ and therefore leads us to an approximation for
$\mathcal{C}$ in a given precision. 
\end{abstract}

\end{frontmatter}


\section{Introduction}
Computing the intersection curve of two surfaces is widely studied in CAGD \cite{Barnhill87,Hoschek93,Patrikalakis02}, which is popularly applied in CAD/CAM and manufacturing.
Existing approaches can typically be classified into numerical and algebraic categories. A very recent method proposed by \cite{Mario06a, Mario06b}, they implemented the typical process to get the numerical intersection curve.   However, some important geometric features of the intersection curve, such as self-intersected points and cusps, might be lost due to their numerical approximation nature.
An efficient algorithm that is robust, accurate, and requiring the least user intervention is therefore needed.

A projectable surface can be regarded as a planar curve in an extended coefficient filed.
We show that projectable surfaces can easily be implicitized
by simple successive resultant computations.
The projectable surfaces include many widely used surfaces, such as whirled surfaces, ruled surfaces and obit-based surfaces. The projectable surface is considerable based on the fact that there are
significant advances on intersection computing for these modeling
surfaces, such as quadrics~\cite{miller95, sarraga83, wang02},
ringed surfaces~\cite{Heo00} and ruled
surfaces~\cite{Mario06a,Heo99,chen11}.

In this paper we compute the intersection curve of two rational parametric surfaces $\S_1(u,s)$ and $\S_2(v,t)$, with $\S_1(u,s)$ being projectable whose implicit form $F(x,y,z)=0$ is computed by our proposed method. By substituting the parametric surface $\S_2(u,s)$ to $F(x,y,z)=0$ we get a plane algebraic curve $G(v,t)=0$. By analyzing the topology graph $\G$ of $G(v,t)=0$ and the singular points on the intersection curve $\mathcal{C}$ we associate a space topology graph to $\mathcal{C}$, which is homeomorphic to $\mathcal{C}$ and therefore leads us to an approximation for
$\mathcal{C}$ in a given precision.

According to the above process, one important work is the implicitization of a given surface.  Many
methods such as the resultant method~\cite{ Mario06a,sed02}, the Groebner basis method~\cite{Cox92} and the $\mu$-basis method~\cite{chen01} were proposed. The resultant method is comparable in the complexity of computation but not complete for general surfaces. The latter two methods are complete but not efficient in practical implements. It is still a problem to implicitize a general surface efficiently. For the
projectable surfaces, we introduce an implicitization method using simple successive resultant computations.
The method is more efficient than others in
these special cases and it can be introduced to numerical
computation.

Another significant work is topology determination of a real algebraic curve. Existing methods of determining the topology are referred to \cite{hong96,gonzalez02,seidel05}.
Based on the concept of segregating box in \cite{chengtop} and real roots isolation of triangular system isolation \cite{chengjsc}, we propose a method to compute the topology graph of a planar algebraic curve inside a box. The given curve need not to be in a generic position.

Some subtle discussions are proposed to refine the topology graph, since there are the points which make the correspondence between the plane curve and the intersection curve not one-to-one. These points are figured out and added to the topology graph. Then each edge of the refined topology graph is homeomorphic to its corresponding curve segment of the intersection curve.

The rest of the paper is organized as follows. In Section 2, we give
some notations. The implicitization method is also proposed for
projectable surfaces. In Section 3 we outline the process to determine the topology graph of a plane curve. In Section 4,  we refine the topology graph and compute the space topology graph of the intersection curve. In Section 5, we give an
algorithm to approximate the intersection curve. Some
experiments are shown in Section 6 and we draw a conclusion in Section 7.

\section{Implicitization of rational projectable surfaces}\label{section2}

A rational parametric surface $\S(u,s)$ is defined by
\begin{equation}
\label{surfacee} (x,y,z)=\S(u,s)=\left(
\frac{p_1(u,s)}{q_1(u,s)},\frac{p_2(u,s)}{q_2(u,s)},
\frac{p_3(u,s)}{q_3(u,s)} \right),
\end{equation} where $p_i,q_i\in \mathbb{Q}[u,s]$ and $\gcd(p_i,q_i)=1, i=1,2,3$.
We assume that the parametrization is proper~\cite{li07} and presents a
non-degenerate surface. Since there still lacks an efficient
implicitization method that applies to general parametric surfaces, we suppose one of the two parametric surfaces, say $\S_1(u,s)$, to be a
\emph{rational projectable surface} that takes the form
\begin{equation}
\label{surf} (x,y,z)=\S(u,s)=\left(
\frac{p_1(u,s)}{q_1(u,s)},\frac{p_2(u,s)}{q_2(u,s)},
\frac{p_3(s)}{q_3(s)} \right).
\end{equation}
The surface~(\ref{surf}) can be treated as a collection of the following planar curves with specified parameter $s$:
\begin{equation}
\label{surfcur} (x,y)=\S(u;s)=\left(
\frac{p_1(u;s)}{q_1(u;s)},\frac{p_2(u;s)}{q_2(u;s)} \right).
\end{equation}
 Since $\gcd(p_i,q_i)=1$ and $\max\{\deg(p_i), \deg(q_i)\} \geq 1,i=1,2$,
\begin{equation}\mathrm{Res}(q_1x-p_1,q_2y-p_2,u)=l(s)L(x,y,s)\label{restmp}
\end{equation} is not identically zero,
where $l(s)\in \mathbb{Q}[s]$ is the content of the resultant w.r.t.
parameters $x,y$. Hence, $L(x,y,s)$ is the primitive part of the
resultant.

If $\mathrm{deg}_s(L)=0$, the surface~(\ref{surf}) is a cylindrical
surface over the $xy$-plane with the irreducible implicit equation $L(x,y)=0$. To determine whether a rational
surface is cylindrical over the coordinate plane is not hard, hence we
consider only the non-degenerate case with $\mathrm{deg}_s(L)\geq 1$. We
have the following theorem.
\begin{theorem}\label{impth}
Let $\S(u,s)$ be a rational projective surface of the form~(\ref{surf}) and its
implicit equation is $F(x,y,z)\not\in \mathbb{Q}[x,y]$. Then up to
a constant multiple
\begin{equation}
\label{imp}F(x,y,z)=\mathrm{Res}(q_3(s)z-p_3(s),L(x,y,s),s).
\end{equation}
\end{theorem}
This theorem is a simplified version of theorem 2 in~\cite{shen08}. This implicitization method is more efficient than most existing approaches. Readers are referred to ~\cite{shen08} for further details.

\subsection{Ruled surfaces}
A ruled surface is formed by a one-parameter family of straight
lines moving along a curve, where the curve is the {\em directrix},
and the straight lines are called {\em rulings}. Ruled surfaces
are widely used in geometric modeling, see \cite{Heo99, Mario06a, chen11} for related work.
Precisely, a rational parametric ruled surface is given by
\begin{equation}\label{eq-r}\begin{array}{l}
\S(u,s)=\left(\frac{a_0(u)+a_1(u)s}{d_1(u)},\frac{b_0(u)+b_1(u)s}{d_2(u)},
\frac{c_0(u)+c_1(u)s}{d_3(u)}\right)\end{array}\end{equation} where $a_{0;1},
b_{0;1}, c_{0;1}, 0\neq d_{1;2;3}\in\Q[u].$ We assume that the
parametric equations in (\ref{eq-r}) are irreducible fractions, for instance,
$\gcd(a_0,a_1,d_0)=1.$
Although the parametrization (\ref{eq-r}) does not agrees with (\ref{surf}), the following lemma allows us to transform (\ref{eq-r}) to a rational projectable surface.
\begin{lemma}\label{repofmu}
Let $\S(u,s)$ be a rational ruled surface of the form (\ref{eq-r}).
Then by a birational parameter transformation, $\S(u,s)$ can be
reparameterized by
$$\begin{array}{l}\bar\S(\u,\s) =
\left(\frac{\a_0(\u)+\a_1(\u)\s}{\d_1(\u)},\frac{\b_0(\u)+\b_1(u)\s}{\d_2(\u)},\s\right),\end{array}$$
 where $\u,\s$ are new parameters and $\a_{0;1}, \b_{0;1},
\c_{0;1}, 0\neq \d_{1;2}\in\Q[\u]$.
\end{lemma}
\begin{proof}
  Since $\S$ is a rational ruled surface, $a_1(u)$, $
b_1(u)$, $c_1(u)$ can not simultaneously be identical to zero. Without loss of  generality, we assume
$c_1(u)\neq 0$. By introducing
\begin{equation}\label{eq-rp}\s=\frac{c_0(u)+c_1(u)s}{d_3(u)},
\u=u, \end{equation}
we get
$$s=\frac{d_3(u)\s-c_0(u)}{c_1(u)},u=\u, $$
which when substituted into (\ref{eq-r}) yields
$$\begin{array}{rl}
&\bar\S(\u,\s)\\
=&\left(\frac{a_0(\u)+a_1(\u)\frac{d_3(\u)\s-c_0(\u)}{c_1(\u)}}{d_1(\u)},
\frac{b_0(\u)+b_1(\u)\frac{d_3(\u)\s-c_0(\u)}{c_1(\u)}}{d_2(\u)} \right.,\left.
\frac{c_0(\u)+c_1(\u)\frac{d_3(\u)\s-c_0(\u)}{c_1(\u)}}{d_3(\u)} \right)\\[0.2cm]
=&\left(\frac{a_0(\u)c_1(\u)-a_1(\u)c_0(\u)+a_1(\u)d_3(\u)\s}{c_1(\u)d_1(\u)}\right.,\left.
\frac{b_0(\u)c_1(\u)-b_1(\u)c_0(\u)+b_1(\u)d_3(\u)\s}{c_1(\u)d_2(\u)},\s
\right)\\[0.2cm]
 =&
\left(\frac{\a_0(\u)+\a_1(\u)\s}{\d_1(\u)},\frac{\b_0(\u)+\b_1(\u)\s}{\d_2(\u)},\s\right).
\end{array}$$
 This gives the parametrization of a projectable surface.
\qed\end{proof}

Once the ruled surface (\ref{eq-r}) is reparameterized by Lemma \ref{repofmu}, we can apply Theorem~\ref{impth} to compute its
implicit equation.

\begin{example}Let $(x,y,z)=\S(u,s)$ be a ruled surface given by
$$\begin{array}{l}\left({\frac {1-{u}^{2}-2\,su}{1+{u}^{2}}},{\frac {2\,u+s(1-{u}^{2})}{1+{u}^{2}}},s \right)
.\end{array}$$
Since the parametrization is already projectable,  we directly apply Theorem \ref{impth}. First we compute
$$L(x,y;s)=4\,{y}^{2}+4\,{s}^{2}{x}^{2}+4\,{s}^{2}{y}^{2}-8\,{s}^{2}-4\,{s}^{4}+4
\,{x}^{2}-4.$$
By removing the content $(4\,{s}^{2}+4)$, i.e.,  the gcd of the
coefficients of $L(x,y;s)$,  we get the primitive part $-{s}^{2}+{y}^{2}-1+{x}^{2}$.  Then the implicit equation of the ruled surface is
$$-{z}^{2}+{y}^{2}-1+{x}^{2}=0.$$
For comparison, readers can see \cite{Mario06a} for the implicitization of the same ruled surface by computing the gcd of three
resultants.
\end{example}

\subsection{Generalized revolution surfaces}
Revolution surfaces are also popularly used in manufacturing, such as porcelain modeling. A rational generalized revolution surface is defined by
$$\begin{array}{l}\S(u,s) = \left(\frac{p_1(s)}{q_1(s)}\frac{2u}{1+u^2},
\frac{p_2(s)}{q_2(s)}\frac{1-u^2}{1+u^2},\frac{p_3(s)}{q_3(s)}\right).
\end{array}$$
When $p_1/q_1=p_2/q_2$, this defines a usual revolution surface rotating around the $z$-axis.
Since this parametrization agrees with the form~(\ref{surf}), we can use Theorem \ref{impth} directly for implicitization.

\subsection{Orbit-based surfaces}
An orbit-based surface is a rational surface formed by translating a
plane curve with its posture unchanged along a space curve. One can find that the orbit-based surface is a special case of the sweep surface.
For instance, a tube surface can be defined by a circle set whose
center follows a space curve
$$\begin{array}{l}
  \S(u,s) = \left(\frac{2u}{1+u^2}-\frac{p_1(s)}{q_1(s)},
\frac{1-u^2}{1+u^2}-\frac{p_2(s)}{q_2(s)},\frac{p_3(s)}{q_3(s)}\right),
\end{array}$$
where $(p_1/q_1,p_2/q_2,p_3/q_3)$ represents a space curve, see
Example~\ref{ex6} in  Section~\ref{experience}.
\begin{remark}
We have shown an efficient approach of implicitizing rational projectable surfaces (up to a birational parameter transformation). Efficiently implicitizing arbitrary rational parametrized surfaces is still left an open problem. Notably
the following process applies to compute the intersection loci of two general surfaces with one rational parametrized and the other being in implicit form.
\end{remark}

\section{Topology determination of planar algebraic curves}\label{section3}
By the method proposed in Section 2, we compute the implicit equation $F(x,y,z)=0$ of the projectable surface $\S_1$. Substituting $\S_2(u,t)$ to $F(x,y,z)=0$ yields a plane algebraic curve $G(v,t)=0$. We next determine the topology graph $\G$ of the curve $G(v,t)=0$ inside a given rectangle.

There are many related work about computing the topology of algebraic curves \cite{chengsocg,eme2011,gonzalez02,hong96}.
We prefer the methods which need not require the curve to be in a generic position and need not to compute a Sturm-like polynomial sequence. We use the concept of segregating boxes in \cite{chengtop} to determine the adjacency relationship when we compute the topology of algebraic curves and real roots isolation of triangular system \cite{chengjsc} to get the critical points of the curve.
We will compute the topology of the curve inside a bounding box.

\begin{definition}
A point $P_0=(v_0, t_0)$ is said to be a \emph{singular point} on the curve
$G(v,t)=0$ if $G(v_0, t_0)=G_v(v_0, t_0)=G_t(v_0, t_0)=0$.
A point $P_0=(v_0, t_0)$ is said to be an $v$-critical point
(resp. $t$-critical point) of $\C$ if $G(v_0, t_0)=0$ and
$G_t(v_0, t_0)=0$ (resp. $G_v(v_0, t_0)=0$).
\end{definition}

\begin{definition}
Let $P$ be a point on the curve $G(v,t)=0$.  The {\em left (right) branch
number of $P$} is the number
of curve segments of $\C$ that passes through $P$ from the left(right) in a small neighbor of $P$.
\end{definition}

The following definition is taken from \cite{chengtop}.
\begin{definition}\ Let $f(x,y)\in\Q[x,y]$ be the defining polynomial of an algebraic curve and $g(x)$ its discriminant with respect to $y$. A {\em Segregating box} of an $x$-critical point $P: (\alpha,\beta)$  of an algebraic curve $f(x,y)=0$ such that $g(\alpha)=f(\alpha,\beta)=0$ is a rectangle $[a,b]\times[c,d]$ containing $P$ inside such that
\begin{enumerate}
\item There is no real roots of $g(x)=0$ in $[a,b]$.
\item There is no real roots of $g(x)=f(x,y)=0$ in $[a,b]\times[c,d]$.
\item The upper and bottom boundaries of $[a,b]\times[c,d]$ have no intersection with the curve $f(x,y)=0$.
\end{enumerate}

\end{definition}

Suppose $G(v,t)$ is square free and contains no univariate factor(s) in $v$\footnote{
The topology of a given curve $\mathcal{C}: G(v,t)=0$ is the same as that of the curve defined by the square free part of
$G(v,t)$. Moreover, if $\mathcal{C}$ contains vertical line(s), these vertical lines can later be added
after the main part of
$\mathcal{C}$ is analyzed. }.

We will compute the topology of curve inside a box $\mathbf{B}=[A,B]\times[C,D]\subset\mathbb{Q}^2$,
and then determine the
{\em topology graph} $\G= \{\PP, \PE\}$, where $\PP$ and
$\PE$ inside the box are defined as follows.
\begin{itemize}
\item $\PP$ is a set of points in the $v-t$ plane:
\begin{equation}\label{eq-ps2} \PP = \{P_{i, j}=(\alpha_i,  \beta_{i, j}),  0\le i\le s,  0\le j \le s_i
\}\end{equation}
where $s,  s_i\in\N$ and $(\alpha_i, \beta_{i, j})$ are towers of
real algebraic numbers such that $\alpha_0<\alpha_1<\cdots<\alpha_s$
and $\beta_{i, 0}<\beta_{i, 1}<\cdots<\beta_{i, s_i}$.
The points $P_{i,j}$ shall later be solved from the
triangular systems $\Sigma_i= \{h_i(v), g_i(v,t)\}$ and then represented by the isolation
boxes $B_{i,j} = [a_i,b_i]\times[c_{i, j},d_{i, j}]$.
Note that  $\G$ or $\PP$ has $s+1$ {\em columns of points}.

\item $\PE=\{(P_1, P_2) | P_1, P_2\in\PP, $ s.t. either
$P_1=P_{i, p}$, $P_2=P_{i+1, q}$ \hbox{ or } $P_1=P_{i, p}, P_2=P_{i,
p+1}\}$.
In the first case,  the edge is called {\em non-vertical}, while in the
second case, the edge is called {\em vertical}. We shall further assume
no intersection between any two edges except at the endpoints.

\end{itemize}

%

The following process outlines our approach to computing the topology graph $\G$:
\begin{alg} Compute the topology of a planar algebraic curve $G(v,t)=0$ inside a bounding box $\mathbf{B}=[A,B]\times[C,D]$.
\end{alg}

\begin{description}
  \item[Step 1] Compute $d(v):=(v-A)(v-B)G(v,C)G(v,D)\mathrm{Res}(G,\frac{\partial G}{\partial t},t)$.

  \item[Step 2] Solve for the real roots of the triangular system
$\Sigma=\{d(v),G(v,t)\}=0$ by the real root isolation method given in \cite{chengjsc}. We take only the real roots inside $\mathbf{B}$.
  \item[Step 3] For each critical point $P=(\alpha_i, \beta_{ij})$ solved from Step 2, construct a segregating box $[a_i,b_i]\times
[c_{i,j},d_{i,j}]$. The number of the left branches
of $P$ is the number of roots of $G(a_i,t)=0$ inside the
interval $[c_{i,j},d_{i,j}]$. Note that the line segments $\{a_i\le v\le b_i,
t=c_{i,j},d_{i,j}\}$ have no intersections with the given curve. The number of right branches for each critical points is similarly computed.
 \item[Step 4] Construct a topology graph of $G=0$ inside $\mathbf{B}$.
\end{description}

\section{Topology determination of the intersection curve $\mathcal{C}$}\label{section4}

\emph{Geometric character points} on the surfaces $\S_1$ and $\S_2$, such as ordinary irregular points, cusps and
self-intersected points, are crucial to determining the space topology graph homeomorphic to the space
intersection curve $\mathcal{C}$.

\begin{definition} Let $\S(v,t)$ be a rational parametric surface. A parameter $(v_0,t_0)$ is said to be regular if $\frac{\partial \S}{\partial v}$ and $\frac{\partial \S}{\partial t}$ are linearly independent at $(v_0,t_0)$; otherwise $(v_0,t_0)$ is said to be irregular.
\end{definition}

\begin{definition} Let $F(x,y,z)=0$ be an implicit surface (of a parametric surface $\S(v,t)$). A point $Q=(x_0,y_0,z_0)$ is a singular point on $F(x,y,z)=0$ if $x_0,y_0,z_0$ is a solution for
$F=\frac{\partial F}{\partial x}=\frac{\partial F}{\partial y}=\frac{\partial F}{\partial z}=0$; otherwise $Q$ is said to be non-singular.
\end{definition}

Suppose a projectable surface $\S_1$ (up to a birational parameter transformation) and a
rational surface $\S_2$ are given by
\begin{equation}\label{intequs}
  \begin{array}{l}\displaystyle \S_1(u,s)=\left(
\frac{f_1(u,s)}{g_1(u,s)},\frac{f_2(u,s)}{g_2(u,s)},
\frac{f_3(s)}{g_3(s)} \right),\\[0.2cm]
\displaystyle\S_2(v,t)=\left(
\frac{p_1(v,t)}{q_1(v,t)},\frac{p_2(v,t)}{q_2(v,t)},
\frac{p_3(v,t)}{q_3(v,t)} \right),
\end{array}
\end{equation}
and $F(x,y,z)=0$ is the implicit equation of $\S_1$ computed by
Theorem~\ref{impth}\footnote{During implicitization, there might be some real
points on $F(x,y,z)=0$ but not on the
parametric surface $\S_1(u,s)$. A connected component formed by such
points is called a \emph{geometric extraneous
component}~\cite{Mario05,Mario06b}. Such points are singular points of the surface
$F(x,y,z)=0$. As mentioned in~\cite{Mario06b}, we need to check singular points in order to verify if they
belong to a geometric extraneous component, which must be removed.}. If a point $\S_2(v,t)$ lies on the intersection curve $\mathcal{C}$, the parameter $(v,t)$ must satisfy
 $F(\S_2(v,t))=0$. Let
$$\tilde G(v,t):=F(p_1(v,t)/q_1(v,t),p_2(v,t)/q_2(v,t),p_3(v,t)/q_3(v,t)),$$
and define the square-free part of the numerator by
$$G(v,t)=\mathrm{Sqf}(\mathrm{numer}(\tilde G(v,t))=0.$$
The intersection curve $\mathcal{C}$ of the surfaces $\S_1$ and $\S_2$ is then determined by
\begin{equation}   \label{inter}
\mathcal{C}^I:\left\{\begin{array}{l}
 G(v,t)=0\\
  x=p_1(v,t)/q_1(v,t)\\
   y=p_2(v,t)/q_2(v,t)\\
    z=p_3(v,t)/q_3(v,t).
\end{array}\right.
\end{equation}
Based on the topology of plane curve  $G(v,t)=0$, one can divide the
intersection curve $\C^I$ into different curve segments. 
We now consider the parameter correspondence between the plane curve
$G(v,t)=0$ and the intersection curve $\C^I$ of the two surfaces.

\subsection{Self-intersection points}\label{section4-1}

\begin{lemma}\label{corres}
Except for the singular points on the plane curve $G(v,t)=0$ and the irregular parameters for the surface $\S_2$,
the tangent of the point on the intersection curve~\eqref{inter} is different from zero.

\end{lemma}
\begin{proof}

In the neighborhood of a point $(v,t)$, one can regard $v$ is the function of $t$, i.e., $v=v(t)$ such that $G(v(t),t)=0$, and $\S_2(v,t)=\S_2(v(t),t)$. By Implicit Function Theorem,  one has $\frac{\partial v}{\partial t}=-\frac{\partial G}{\partial t}/\frac{\partial G}{\partial v}$.
Hence the tangent vector ${\mathbf w}$ to the intersection curve $\mathcal{C}^I$ at the point $\S_2(v,t)$ is
\begin{equation*}
\begin{aligned}
{\mathbf w}=\frac{\partial \S_2(v(t),t)}{\partial t}&=\frac{\partial \S_2}{\partial v}\frac{\partial v(t)}{\partial t}+\frac{\partial \S_2}{\partial t}.
\end{aligned}
\end{equation*}
Since, then up to a constant multiple
$$\mathbf w=\frac{\partial \S_2}{\partial v}\frac{\partial G}{\partial t}-\frac{\partial \S_2}{\partial t}\frac{\partial G}{\partial v}.$$
For a parameter pair $(v_0,t_0)$, if neither $\frac{\partial \S_2}{\partial v} \times \frac{\partial \S_2}{\partial t}$ nor $(\frac{\partial G}{\partial v}, \frac{\partial G}{\partial t})$ vanish at $(v_0,t_0)$, i.e., the parameter $(v_0,t_0)$ is regular on the surface $\S_2$ and the point $(v_0,t_0)$ is non-singular on the plane curve $G(v,t)=0$, we have $\mathbf w\not=\mathbf{0}$, which means that the tangent of the point $\S_2(v_0,t_0)$ is different from zero on the intersection curve $\C^I$.
\qed\end{proof}

According to Lemma~\ref{corres}, besides those singular points $(v,t)$ on the plane curve $G(v,t)=0$ (which are already computed in Section 3),  the
irregular points on the surface $\S_2(v,t)$ shall also lead to singular points on the intersection curve $\C^I$. These irregular points $(v,t)$ are solutions for
\begin{equation}
\label{irrgular}
\left\{
\begin{array}{l}
\frac{\partial \S_2}{\partial v}\times \frac{\partial \S_2}{\partial
t}=\mathbf{0}\\
G(v,t)=0.
\end{array}\right.
\end{equation}
A bad situation may occur for~\eqref{irrgular}, that is, the surfaces $\S_2$ has a irregular parameter locus that corresponds to points on the intersection curve $\mathcal{C}^I$. Under this situation, \eqref{irrgular} has an infinite number of solutions. We shall first remove this common irregular parameter locus which shall be treated as a
special curve component.

As known that a point $Q$ is a \emph{cusp} of the $\mathcal{C}^I$ if the tangent of $Q$ is vanish. Hence, the parameters corresponding to the cusp are included in the singular points of $G(v,t)=0$  and the solutions of \eqref{irrgular}.

Notably, some self-intersection points of the intersection curve $\C^I$ may neither correspond to singular points on the plane curve $G(v,t)=0$ nor correspond to
the irregular parameters on the surface $\S_2(v,t)$. See the following example.
\begin{example}\label{addpoint}Given two surfaces
$$\S_1:\left\{\begin{array}{l}
x=(s u^2+s+1)/2\\
y=su\\
z=u
\end{array}\right.\, \,
\S_2:\left\{\begin{array}{l}
  x=v\\
   y=\left( t-1 \right) t-1/4\,v\\
    z= \left( t+1 \right)  \left( t-1
 \right) t
\end{array}\right.,$$
where $\S_1$ is an elliptic paraboloid with whose
implicit equation is $2x-y^2-z^2-z=0$. By computation
$G(v,t)=2\,v+{t}^{4}+{t}^{3}+{t}^{2}v/2-2\,{t}^{2}-tv/2-{v}^{2}/16-{t
}^{6}+t=0$, which has no singular points.
\begin{figure}[!h]
 \centering
 \includegraphics[width=0.31\textwidth,height=2.1in]{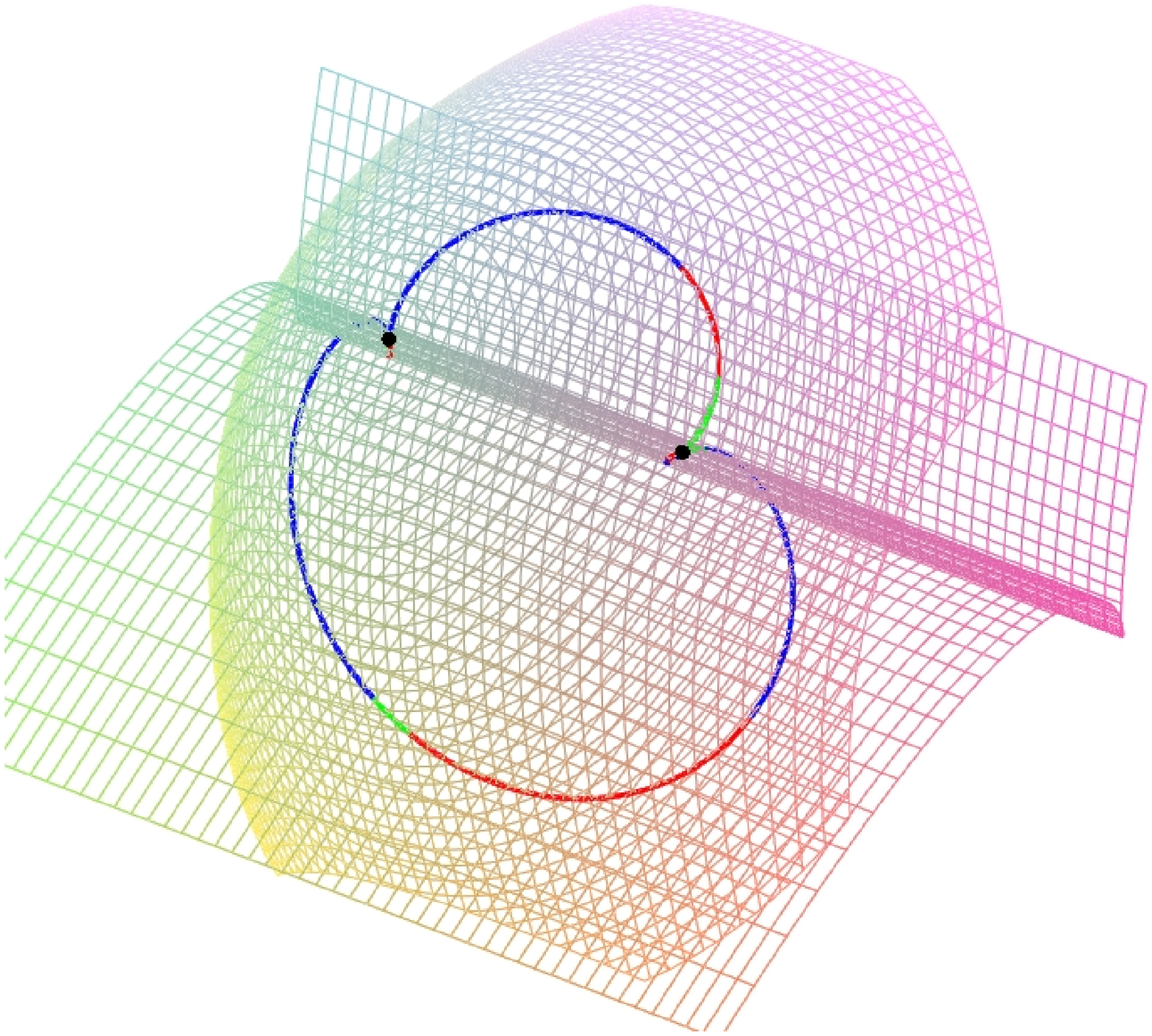}\
  \includegraphics[width=0.33\textwidth,height=2.1in]{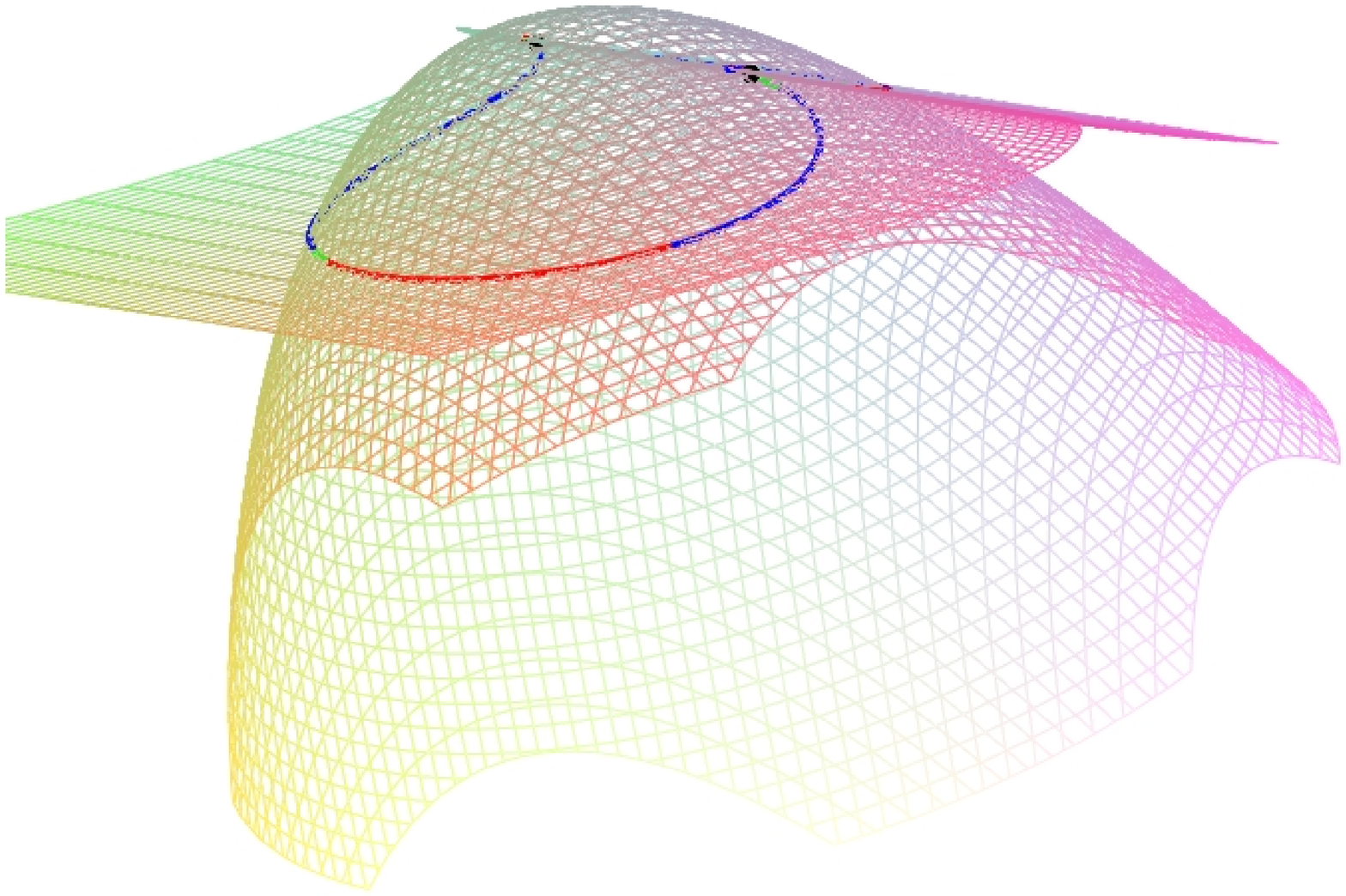}
  \caption{Intersection curve of $\S_1$ and $\S_2$} \label{totalint}
\end{figure}
However, the intersection curve $\C^I$ has two self-intersection points $(0,0,0)$ and $(32,-8,0)$ whose corresponding
parameters on $G(v,t)=0$ are $\{(0,0),(0,1)\}$ and $\{(32,0),(32,1)\}$
respectively~(Figure~\ref{totalint}). One can check that these two
points are regular in $\S_2(v,t)$.
\end{example}

\begin{figure}[!h]
 \centering
 \includegraphics[width=0.22\textwidth,height=1.2in]{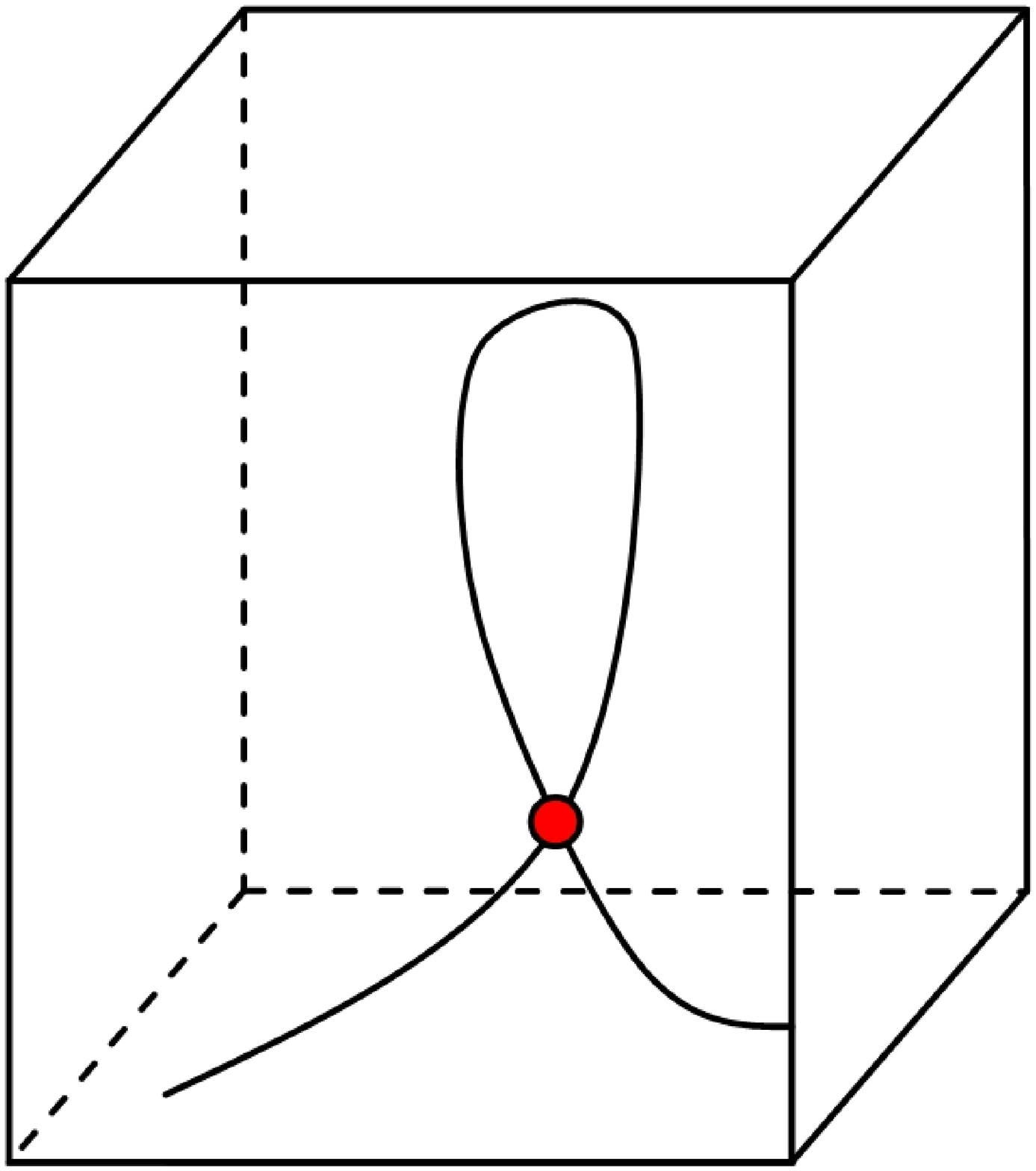}
 \includegraphics[width=0.2086\textwidth,height=1.2in]{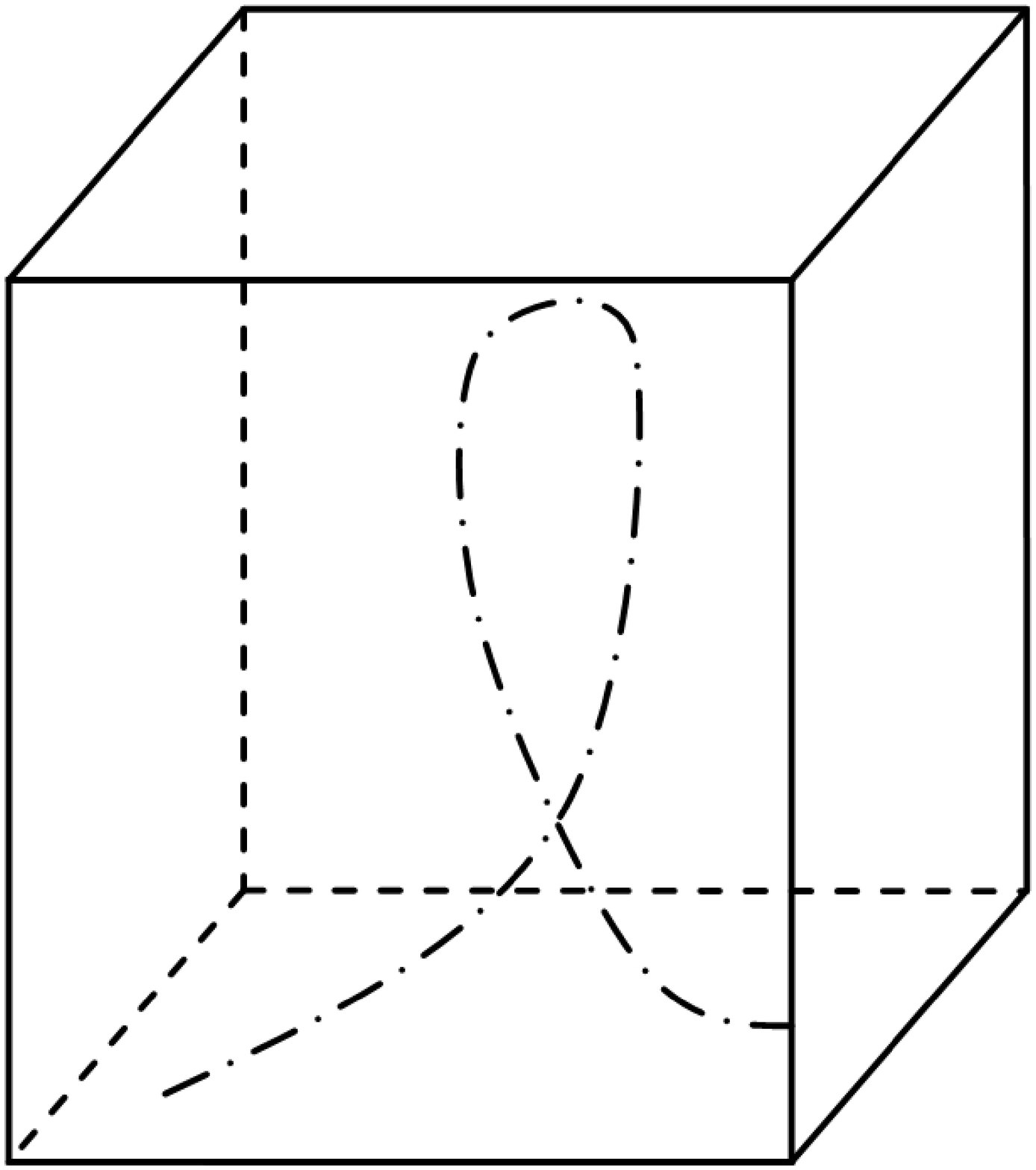}
 \caption{Self-intersection point overlooked by numerical approximation} \label{pointmiss}
\end{figure}

This subtle situation did not draw enough attention in previous work, such as in ~\cite{Mario06a}. This might omit some
self-intersection points on the curve $\C^I$, whose corresponding parameters are both non-singular points  on $G(v,t)=0$ and regular points on surface $\S_2(v,t)$. See Figure \ref{pointmiss}.
Therefore, to ensure that the associated space graph $\G^I$ mapped from $\G$ is
homeomorphic to the intersection curve $\C^I$, here we refine the topology graph $\G$ by adding the following character points of the intersection curve $\mathcal{C}^I$:
\begin{equation}
\label{E:addvertice}
\{P_{i, j}=(\alpha_i, \beta_{i, j})\big|\big(x(\alpha_i, \beta_{i,
j}), y(\alpha_i, \beta_{i, j}),z(\alpha_i,  \beta_{i, j} )\big)\text{~is a self-intersection point on }\C^I\}
\end{equation}
 with $0\le
i\le s, 0\le j \le s_i$.
Note that this might reintroduce some points $(v,t)$ that are already computed in Section 3. Our principal here is not to omit any possible $(v,t)$ that is crucial to the topology of the space curve $\mathcal{C}$ but do not pursue the complement of the previous computed $(v,t)$.

We now show how to compute the self-intersection points of~\eqref{inter}.
The self-intersection points $(x,y,z)$ of $\mathcal{C}^I$ are those self-intersection points on $\S_2(v,t)$ that also lie on $\S_1$, and hence can be solved through:
\begin{equation}
\left\{ \begin{array}{l}
x(v,t)-x(u,s)=0\\
y(v,t)-y(u,s)=0\\
z(v,t)-z(u,s)=0\\
G(v,t)=0,G(u,s)=0.
\end{array}\right.
\end{equation}
Removing the denominators yields
\begin{equation}\label{self-int}
\left\{ \begin{array}{l}
p_1(v,t)q_1(u,s)-p_1(u,s)q_1(v,t)=0\\
p_2(v,t)q_2(u,s)-p_2(u,s)q_2(v,t)=0\\
p_3(v,t)q_3(u,s)-p_3(u,s)q_3(v,t)=0\\
G(v,t)=0, G(u,s)=0.
\end{array}\right.
\end{equation}
Except for the trivial solution set $\{v=u,t=s\}$, the
system~\eqref{self-int} can only has a finite number of solutions which
correspond to the self-intersection points of the intersection curve $\C^I$. These solutions can be obtained by
Ritt-Wu's characteristic set method~\cite{wu84}, supported by the maple packages \texttt{Wslove} and \texttt{Charsets} by D.K.Wang and
D.M. Wang respectively. The packages can be downloaded from \url{
http://www.mmrc.iss.ac.cn/\~dwang/soft.htm} and
 \url{http://www-calfor.lip6.fr/\~wang/epsilon/}.

Notably during the zero decomposition in Ritt-Wu's method, some multiplicities of the solutions may be
  lost. However, we do not care about these multiplicities, since the multiple solutions are corresponding to the cusp points, which are already
  computed.

\begin{theorem}\label{one2one}
Except for the singular points on the plane curve $G(v,t)=0$, the irregular parameters of the surface $\S_2$ from~\eqref{irrgular} and the parameters of self-intersection points from~\eqref{self-int}, there is a one-to-one
correspondence between the plane points on the plane curve $G(v,t)=0$ and the space points on the
intersection curve $\C^I$ of the two surfaces.
\end{theorem}
\begin{proof}
By Lemma~\ref{corres}, the tangent is well defined at the space point on the
intersection curve $\C^I$ of the two surfaces. Furthermore, the self-intersection points are excluded in equation system~\eqref{self-int}. Hence, except for these points, there is a one
to one correspondence between the plane curve and the intersection
curve of the two surfaces.
\qed\end{proof}

\subsection{Space topology graph}
We now determine the space topology graph $\G^I$, whose vertices are mapped from the vertices of the refined graph of
$\G$: $$\{P_{i, j}^I=((x(\alpha_i, \beta_{i,
j}),y(\alpha_i, \beta_{i, j}),z(\alpha_i,  \beta_{i, j})):\alpha_i,
\beta_{i, j})\},~~~0\le i\le s, 0\le j \le s_i,$$ and edges
$(P_{i_1,j_1}^I,P_{i_2,j_2}^I)$ represent the topological connections between the points
$P_{i_1,j_1}^I$ and $P_{i_2,j_2}^I$ on $\mathcal{C}^I$. Note that any two edges $(P_i^I,P_j^I)$ and $(P_k^I,P_l^I)$ can only share one of their endpoints.

\begin{alg}\label{spacegraph}Compute the refined topology graph $\G$ of the irreducible curve $G(v,t)=0$ and the space topology graph $\G^I$ of the space curve~\eqref{inter}.
\begin{enumerate}
  \item Compute the critical and singular points of $G(v,t)=0$
  and determine the topology graph $\G$ by Section~\ref{section3}.
  \item Compute the cusp and self-intersected points
  of the space curve~\eqref{inter} by the methods in Section~\ref{section4-1}.
  \item Refine the the topology graph $\G$ by adding the parameters corresponding to the points computed in step 2, as well as the edges.
  \item Map the vertices of the refined graph of $\G$ to space points to get the vertices of the space graph $\G^I$, and correspondingly map the edges in the refined graph of $\G$ to the edges of the space graph $\G^I$.
   \item   If any pair of the edges $(P_i^I,P_j^I)$ and $(P_k^I,P_l^I)$ in $\G^I$
  have no intersection points except at the endpoints, then output $\G$ and $\G^I$. Otherwise, if two edges $(P_i^I,P_j^I)$ and $(P_k^I,P_l^I)$ intersects at transversally, add subdivision vertices $P_{1_m}^\star=(v_{1_m}^\star,t_{1_m}^\star)$, $m=1,\ldots$, $m_0$ to $\G$ between $P_i^I$ and $P_j^I$,
  $P_{2_n}^\star=(v_{2_n}^\star,t_{2_n}^\star),n=1,\ldots,n_0$ between $P_k^I$ and $P_l^I$, such that the line segments $(P_i^I,{P_{1_1}^\star}^I),({P_{1_1}^\star}^I,{P_{1_2}^\star}^I),\ldots$, $({P_{1_{m_0}}^\star}^I,{P_j}^I)$ and $(P_i^I,{P_{2_1}^\star}^I)$, $({P_{2_1}^\star}^I,{P_{2_2}^\star}^I),\ldots$, $({P_{2_{n_0}}^\star}^I$, ${P_j}^I)$ are topology edges of $\G^I$. Go to
  Step 4.
\end{enumerate}
\end{alg}

In Step 4 and 5, one can determine the intersection of two line segments $(P_i^I,P_j^I)$ and $(P_k^I,P_l^I)$ using the bracket formulas in~\cite{chen11}.

\begin{lemma}\label{isotopo}In Algorithm~\ref{spacegraph},
  the output space topology graph $\G^I$ is homeomorphic to the intersection
  curve $\mathcal{C}^I$.
\end{lemma}
\begin{proof}~For two points $P_i,P_j$ on graph $\G$, their edge $(P_i,P_j)$ corresponds to a curve segment $\widetilde{P_iP_j}$ of
$G(v,t)=0$. Map them to the space topology graph $\G^I$, we get
 $P_i^I,P_j^I$ and their line segment $(P_i^I,P_j^I)$ on $\G^I$. The space curve
 segment $\widetilde{P_i^IP_j^I}$ on $\mathcal{C}^I$ is then equivalent to
 $\widetilde{P_iP_j}$. If there exists no singular points on $\widetilde{P_i P_j}$, then by Theorem~\ref{one2one}, $\widetilde{P_i^I P_j^I}$
is a continuous curve segment.
According to Algorithm~\ref{spacegraph}, there has no cusp or self-intersection points on $\widetilde{P_i^I P_j^I}$ except for the endpoints. Hence $\widetilde{P_i^I P_j^I}$ is homeomorphic to the line segment $({P_i}^I, {P_j}^I)$.

If the edge $({P_i}^I, {P_j}^I)$ intersects with another edge  $({P_k}^I, {P_l}^I)$, one can subdivide these two edges by adding a finite number ($m_0$ and $n_0$) of  points in Step 5. This is based on the facts that a curve segment can be approximated in any precision by line segments, and that $\widetilde{P_i^I P_j^I}$ and $\widetilde{P_k^I P_l^I}$ have no intersection points excepted in endpoints.
  Since any two line segments has no intersection point except in the endpoints, the line segments are then the edges of the space topology graph $\G^I$.
   Then the output $\G^I$ is homeomorphic to the intersection
  curve $\mathcal{C}^I$.\qed
\end{proof}

\begin{remark}
In Algorithm~\ref{spacegraph} and Lemma~\ref{isotopo}, we simplify the discussion by assuming
$G(v,t)$ irreducible. In fact,
the algorithm and lemma can be enhanced for general cases $G(v,t)$
by factorizing $G(v,t)$ to irreducible factors $G_i(v,t)$ and
decomposing the curve to the components in assumed form. However, to
combine these decomposed components, we should compute the common
points of $G_i(v,t)=0$ and add them to each topology graph $\G_i$,
since these intersections may be lost in the numerical
computation.
\end{remark}

\begin{example}Continue with Example~\ref{addpoint}, the topology
graph $\G$ of $G(v,t)=0$ not including the self-intersected points of $\mathcal{C}^I$, its mapped topology graph $\G^I$ and the
numerical intersection are shown in Figure~\ref{wanted}. $\G^I$ and the numerical curve loss the self-intersected points of $\mathcal{C}^I$.
\begin{figure}[!htp]
 \centering
 \includegraphics[width=0.18\textwidth]{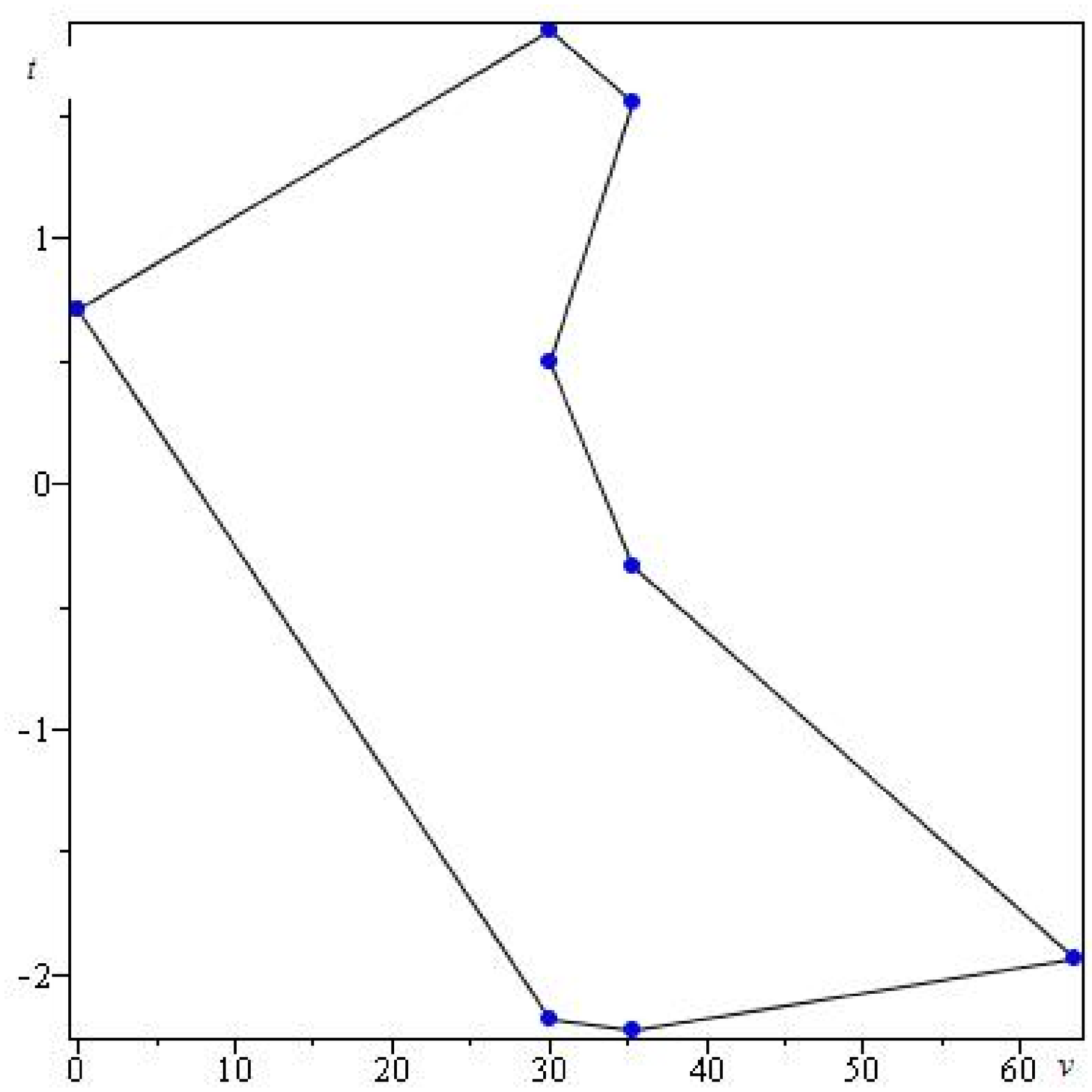}
 \includegraphics[width=0.20\textwidth]{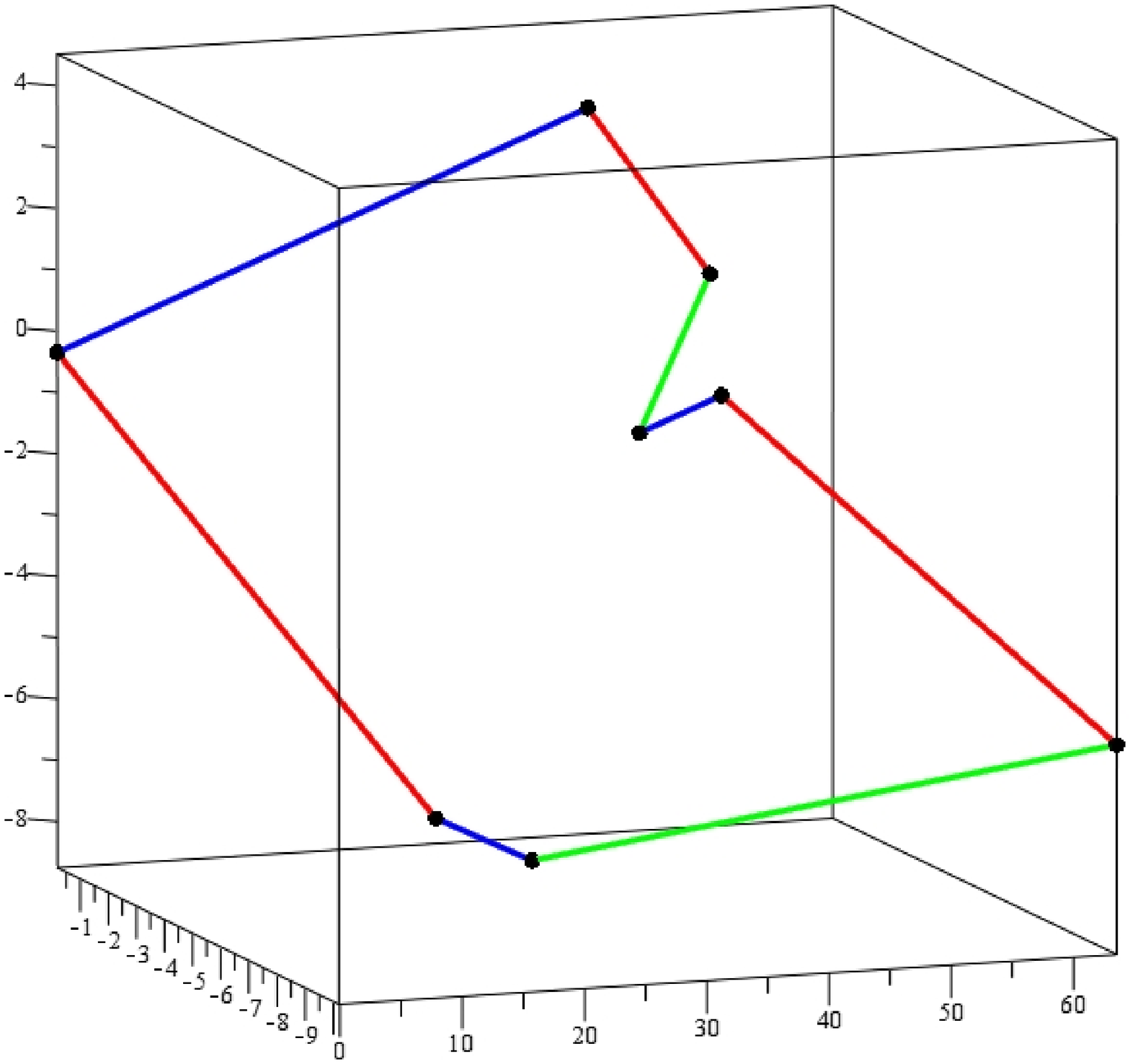}
  \includegraphics[width=0.20\textwidth]{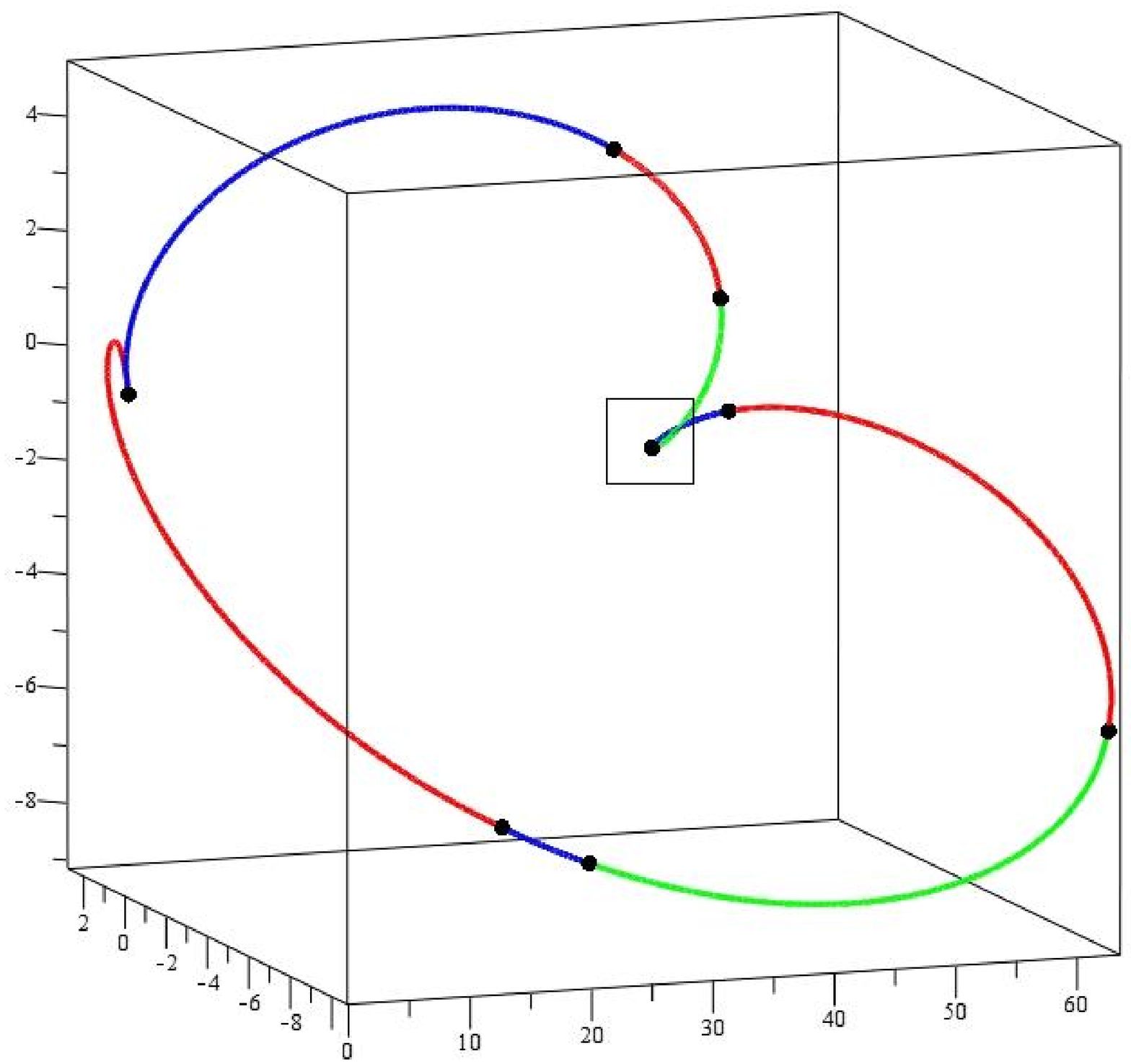}
   \includegraphics[width=0.20\textwidth]{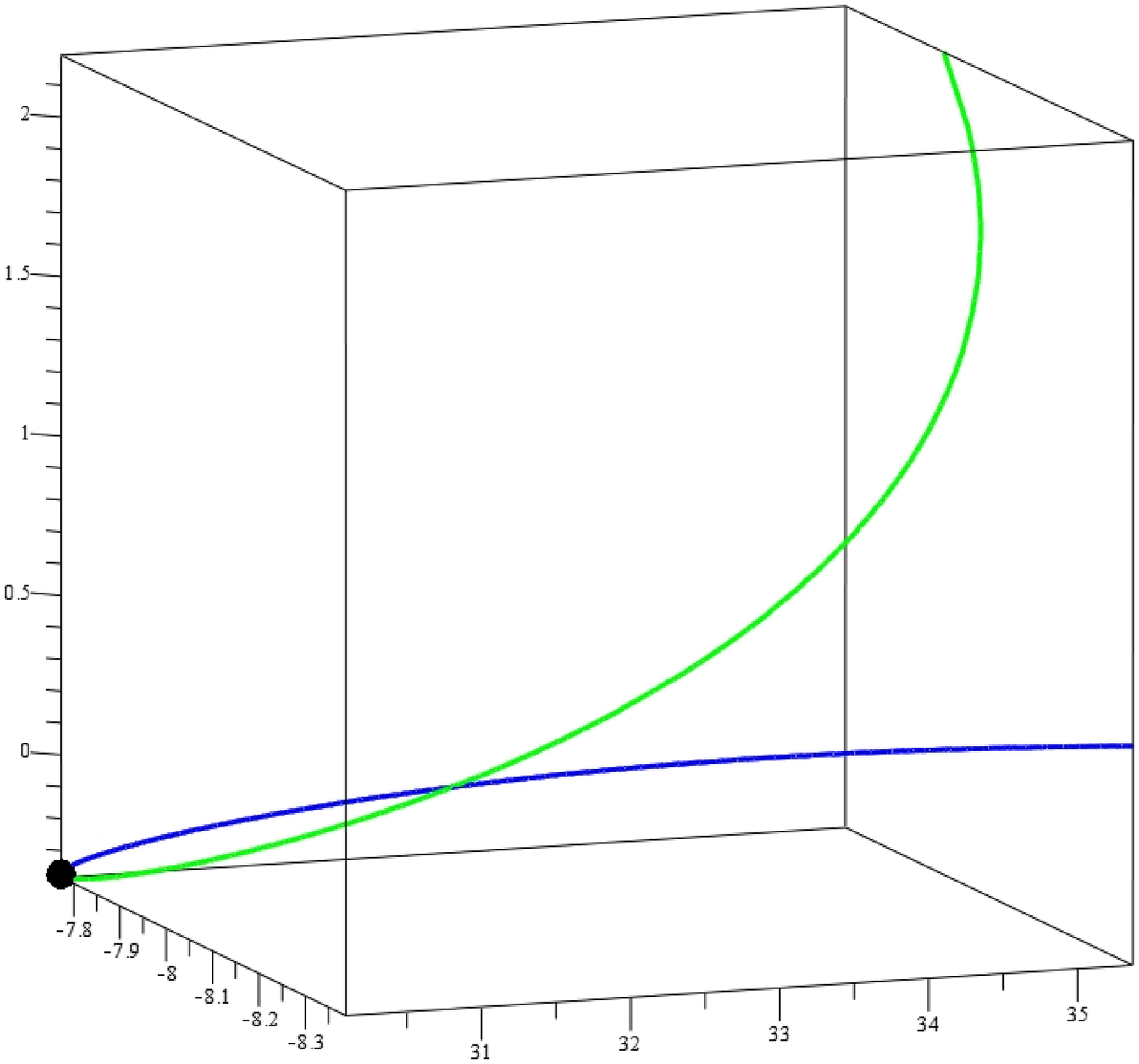}
 \caption{$\G$, $\G^I$, numerical intersection and enlarged neighbor} \label{wanted}
\end{figure}

Adding two self-intersected points, we get the refined $\G$, $\G^I$ and the
numerical intersection as Figure~\ref{added}. The red diamond points
are added since they correspond to the self-intersected points. The refined $\G^I$ and the numerical curve have same topology with $\mathcal{C}^I$.
\begin{figure}[!htp]
 \centering
  \includegraphics[width=0.18\textwidth]{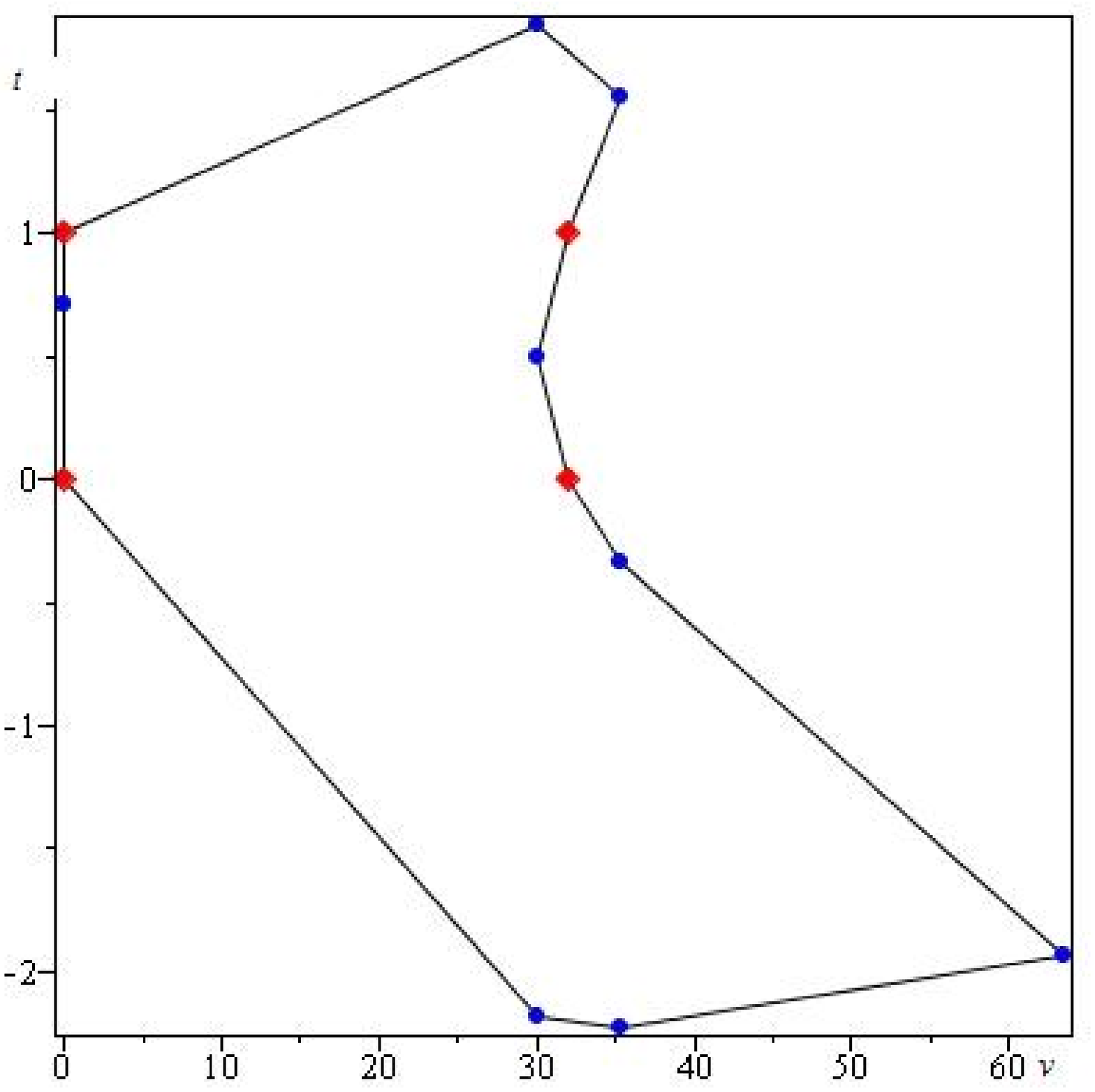}
 \includegraphics[width=0.20\textwidth]{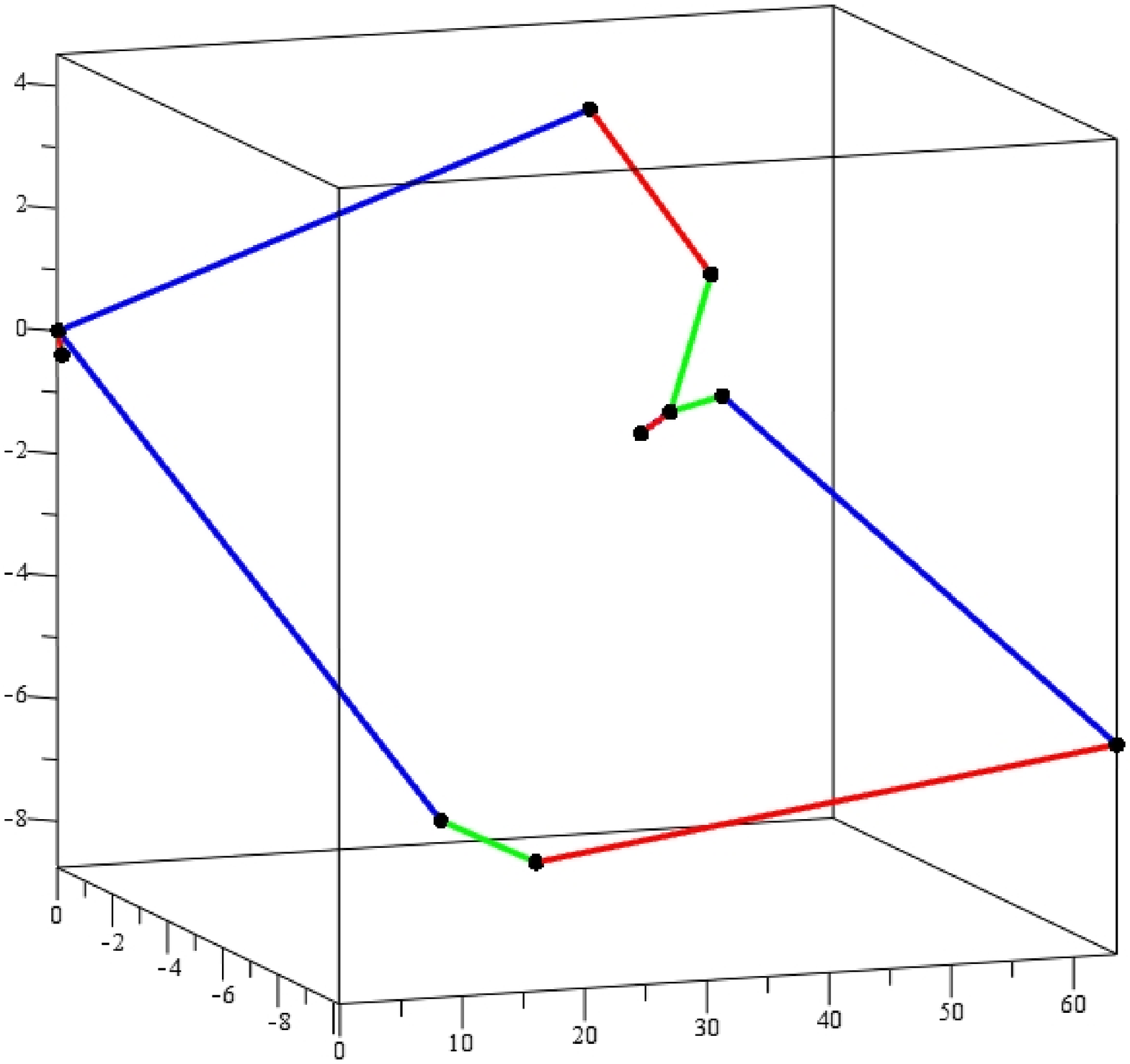}
  \includegraphics[width=0.20\textwidth]{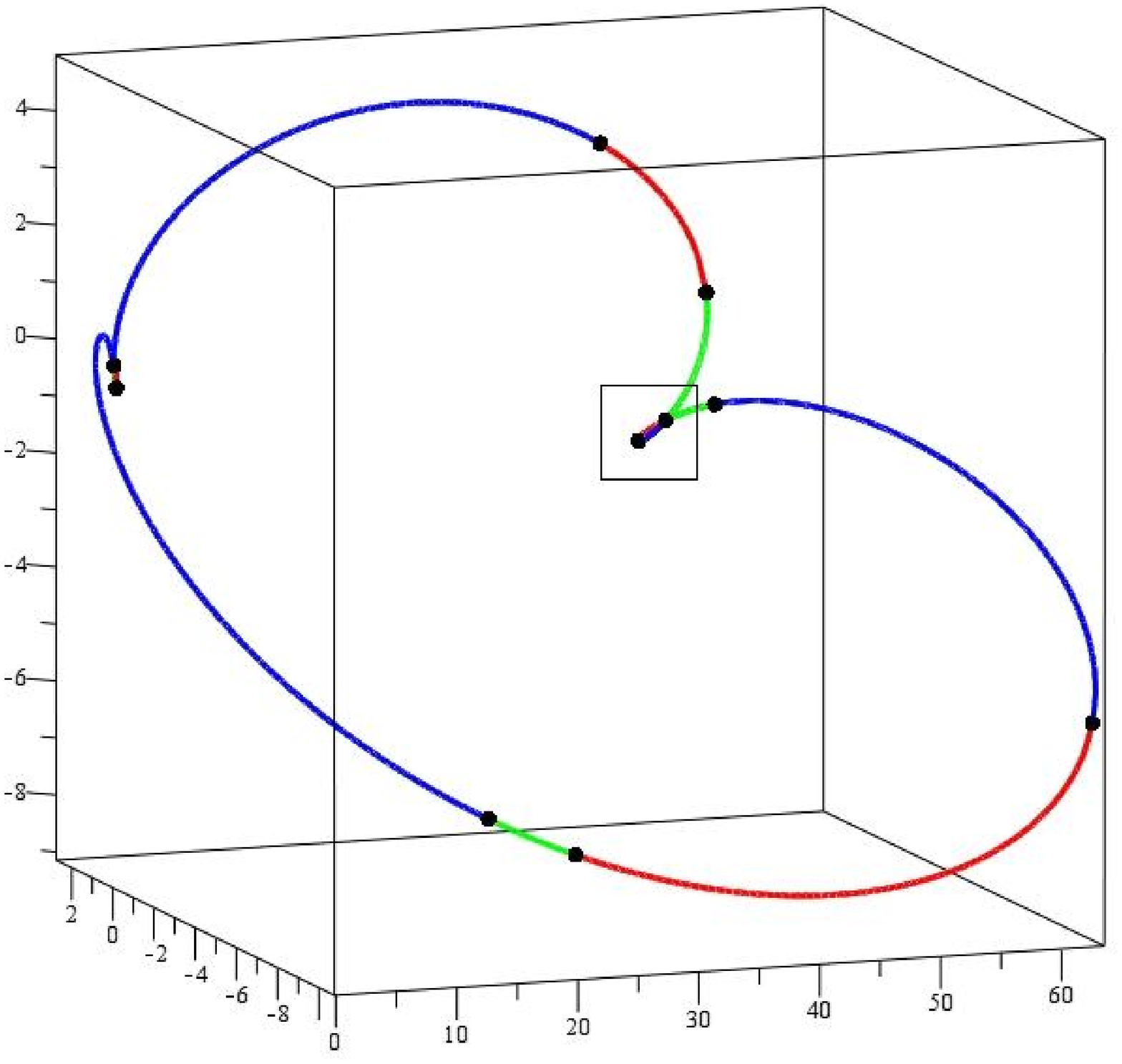}
   \includegraphics[width=0.20\textwidth]{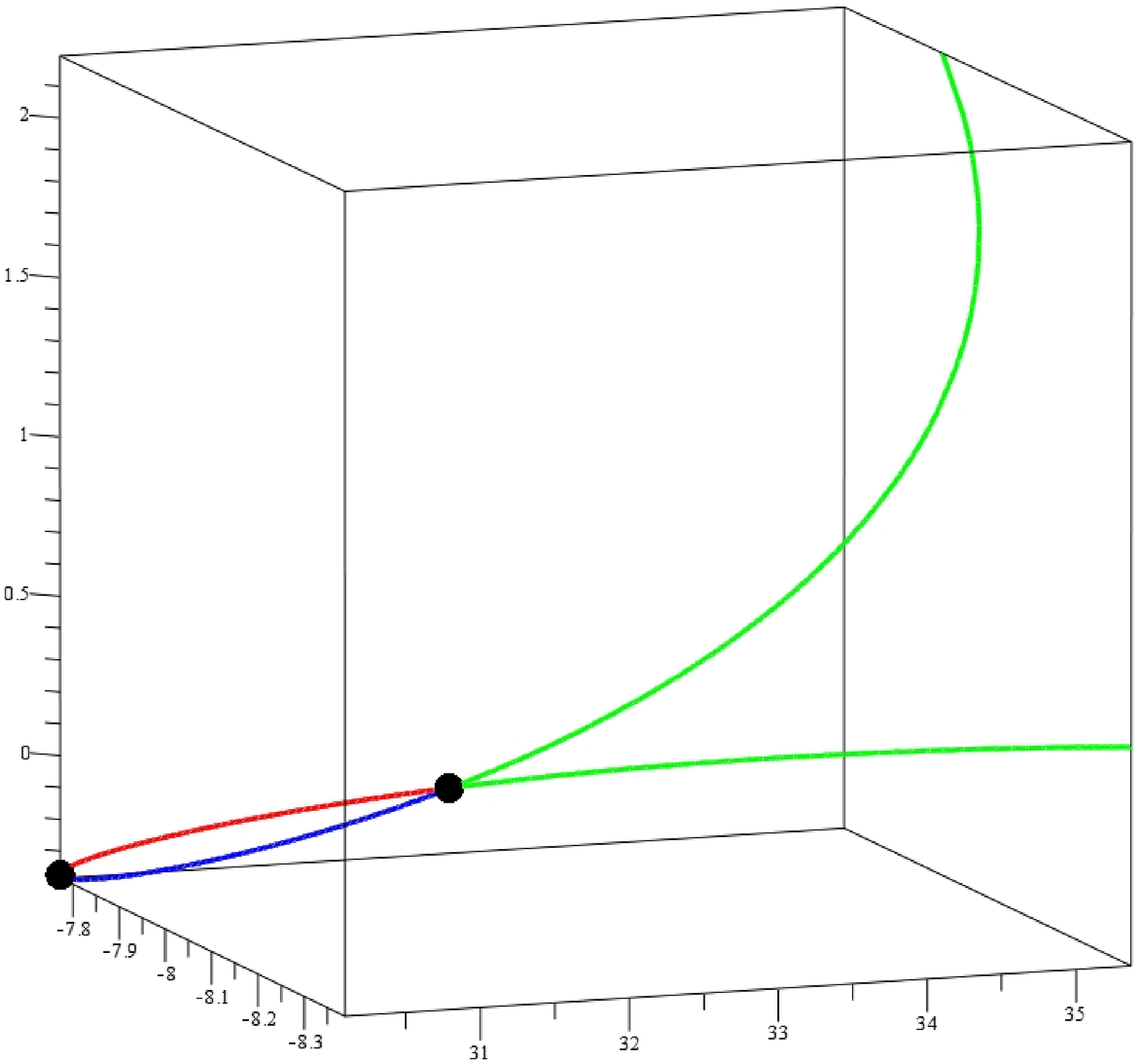}
 \caption{Refined $\G$, $\G^I$, numerical intersection and enlarged neighbor} \label{added}
\end{figure}

The self-intersected points are preserved as a topology vertex in
our numerical intersection now (right one of Figure~\ref{added}) while it may be lost in a numerical approximation (right one of Figure~\ref{wanted}).
\end{example}

\section{Approximation of the intersection curve}
Since the plane topology graph $\G$ and the space topology graph $\G^I$ are both determined in the last section, we now
approximate the intersection curve $\C^I$ within a given precision~$\epsilon$. Our principal is to subdivide the plane topology graph $\G$, and hence the space topology graph $\G^I$ is simultaneously subdivided.

\begin{figure}[!htp]
 \centering
   \includegraphics[width=0.450\textwidth]{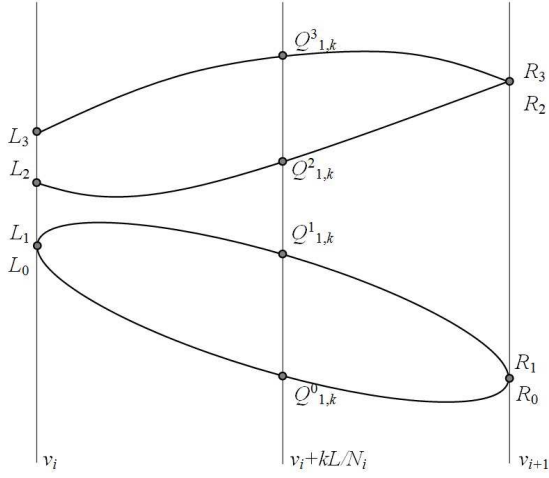}
 \caption{The arrangement of the points in $\mathbf{B}_i=[v_i,v_{i+1}]\times[C,D]$} \label{fig-sub}
\end{figure}

Suppose the vertices of the plane graph $\G$ are
$$\mathcal{P}=\{P_{i,j}(v_i,t_{i,j})\in [a_i,b_i]\times[c_{i,j},d_{i,j}], 0\le i\le s, 0\le j\le s_i\}.$$
Now we consider the boxes $\mathbf{B}_i=[v_i,v_{i+1}]\times[C,D]$.
Suppose there are $m_i$ edges of $\G$ enclosed in $\mathbf{B}_i$, i.e, $m_i$ curve branches originate from the vertices $\{P_{i,j}, 0\le j\le s_i\}$ and end at $\{P_{i+1,j},0\le j\le s_{i+1}\}$. Now for each $i$, rename the vertices of $\G$ on the vertical lines $v=v_i$ and $v=v_{i+1}$ as $\{L_j,0\le j\le m_i-1\}$ and $ \{R_j,0\le j\le m_i-1\}$, respectively. Note that some $x$-critical points may repeat several times in $L$ (or $R$) list. The following procedure tells how to approximate the intersection curve $\C^I$ within a given precision $\epsilon$:

\bigskip
\begin{alg} Approximate $\C^I$ within a given precision $\epsilon$ from the topology of $G(v,t)=0$.
\end{alg}
\begin{enumerate}
\item We first deal with the vertical edges in the graph $\G$ (note that the $v-$ coordinates of these vertical edges are zeros of the content $V(v)$ of $G(v,t)$) if they exist. For a vertical edge $v=v_i$ in $\G$, its corresponding component in the space curve $\C^I$ is $(x(v_i,t),y(v_i,t),z(v_i,t))=\S_2(v_i,t)$.
\item
For $i=0, \ldots,  s-1$, execute the following steps.
\begin{enumerate}
\item Let $L=v_{i+1}-v_i$, and let $N_i$ be the minimal integer larger than $L/\epsilon$; if $N_i=1$, we set $N_i=2$.
Let $t_{i,k}$ be the roots set of
$G(v_i+k \frac{L}{N_i}, t)=0$ inside $[C,D]$, $1 \leq k
\leq N_i-1$. Arrange $t_{i,k}$ from bottom to up, we have
$t_{i,k}=\{t_{i,k}^0,\ldots,t_{i,k}^{m_i-1}\}$.
Note that points $Q_{i,k}^j:=(v_i+k \frac{L}{N_i}, t_{i,k}^j)$ are not $x$-critical points of $\C$.
\item Get the two lists $\{L_j,0\le
j\le m_i-1\}, \{R_j,0\le j\le m_i-1\}$ as mentioned before.
\item For each planar point $L_j, R_j$ or $Q_{i,k}^j$ (sketch shown in Figure \ref{fig-sub}), denoted by $(v_{\alpha},t_{\alpha,\beta})$, compute its
corresponding spatial point $L^I_j, R^I_j$ or ${Q^I}^j_{i,k}$, i.e.,  $\S_2(v_{\alpha},t_{\alpha,\beta})$.
\item For each spatial line segment in : $$\{(L^I_{j}, {Q^I}^j_{i,1}), ({Q^I}^j_{i,k},{Q^I}^j_{i,k+1})_{1\leq k\leq N_i-2}, ({Q^I}^j_{i,N_i-1}, R^I_{j})\}$$
(${j=0, \ldots,m_i-1}$) (assuming the endpoints of the line segment are $P_1(x_1,y_1,z_1)$ and
$P_2(x_2,y_2,z_2)$), check the the Hausdorff distance between the line segment $(P_1^I,P_2^I)$ and its corresponding curve segment $\widetilde{P_1^I P_2^I}$ in $\mathcal{C}^I$. If
\begin{equation}\label{precise}\mathrm{Dis}((P_1^I,P_2^I),\widetilde{P_1^I P_2^I})<\epsilon,\end{equation}
does not hold, subdivide the planar graph $\G$ until all the mapped subdivided
spatial graph $\G^I$ segments satisfy the above condition.
%
\item For any pair of line segments $(P_i^I,P_j^I)$ and $(P_k^I,P_l^I)$, we subdivide them to topology graph edges if they have intersection points.
It means that any pair of edges $(P_i^I,P_j^I)$ and $(P_k^I,P_l^I)$ has no intersection points except for endpoints in the subdivided topology graph.

\end{enumerate}
\end{enumerate}

\begin{remark}
In Step 1, since we use isolation interval for $v_i\in [a_i,b_i],v_{i+1}\in[a_{i+1},b_{i+1}]$, we let $L=(a_{i+1}+b_{i+1}-a_{i}-b_{i})/2$ in practical computation. If $(a_i+b_i)/2+L/N_i<b_i$ (or
$(a_{i+1}+b_{i+1})/2-L/N_i>a_i$), refine $[a_i,b_i]$ (or
$[a_{i+1},b_{i+1}]$).
 \end{remark}

Consider the planar curve segment subdivided in $vt$-plane,
all the endpoints of the segments are the topology vertices on the curve
$G(v,t)=0$, and the endpoints of the corresponding spatial curve
segments are the topology vertices on the intersection curve.
It is clear that the approximation precision between the intersection curve and the numerical approximation
is less than $\epsilon$ since all the curve segments satisfy~\eqref{precise}.

The exact Hausdorff distance in~\eqref{precise} is not easy to compute, we actually compute the numerical distance instead. Choosing $m$ points from $\widetilde{P_1^I P_2^I}$, we then compute the max distance between these points and $(P_1^I, P_2^I)$ as the distance of~\eqref{precise}.

\begin{theorem}
The numerical intersection is homeomorphic to the intersection curve
$\mathcal{C}^I$ and convergence to $\mathcal{C}^I$ in parameters
subdivision process based on $\G$.
\end{theorem}
\begin{proof}
Since the numerical computation is a parameter subdivision process
based on the refined topology graph $\G$, by step 2(e), it is actually a
subdivision of the space topology graph $\G^I$. According to
Lemma~\ref{isotopo}, the numerical intersection is homeomorphic to
$\mathcal{C}^I$.

Since all the character points are computed as vertices, they are
preserved in the parameter subdivision process. We can find that the
numerical intersection curve converges to $\mathcal{C}^I$ as the line
segments approximation.
\qed\end{proof}

The above numerical intersection is a line segment approximation. For further consideration, we can
 give the B-spline approximate
intersection. A method is proposed in~\cite{shen11} to approximate a give space curve based on its topology graph $\G^I$. For each vertex of $\G^I$,
we can compute the left and right tangent directions as well as the
osculating planes. On an ordinary point, the left and right tangents
and normal direction are consistent as well as the osculating
planes. Consider a space curve segment $\widetilde{P_1^IP_2^I}$ with
the tangent direction and osculating planes at endpoints, we can
construct the cubic Bezier curves to approximate $\widetilde{P_1^IP_2^I}$. Then rewrite the Bezier spline curve to B-spline curve with
proper knots selection.

Comparing with the line segment approximation, there are at least three advantages in cubic B-spline approximation. The first one is that the approximation B-spline is $C^1$ continues except in the cusps while the line segment approximation is only $C^0$ continues. Then second is that the cusps are preserved. Finally, the number of approximate curves segments is much less than that in line approximation.

\section{Experiments}\label{experience}

To illustrate our algorithm, we will give some examples in this
section. Some of them are taken form~\cite{Heo99}
and~\cite{Mario06a} for comparison.

\begin{example}Consider the intersection of a cone and an elliptic
cylinder~\cite{Heo99,Mario06a}
$$\begin{array}{l} \S_1=\left({\frac {1-{u}^{2}+s ( 1-{u}^{2}) }{1+{u}^{2}}},{\frac {2
\,u+2\,su}{1+{u}^{2}}},1+s
\right),\\[0.2cm]
 \S_2=\left({\frac {1-{v}^{2}}{1+{v}^{2}}},{\frac {2\,v+t (
1+{v}^{2}
 ) }{1+{v}^{2}}},1+t
\right).
\end{array}$$
Since $\S_1$ is projectable, one can compute its implicit equation and get the
$(v,t)$-plane curve equation
$$G(v,t)=t(1+v^2)(v-1)=0.$$
The solution of line $t=0$ corresponds to the red circle in
Fig.~\ref{figofce}, which is the directrix of both surfaces. The
line $v=1$ corresponds to the common ruling (blue line
in Fig.~\ref{figofce}) where the two surfaces  meet tangentially.
The intersection point of the circle and the line  is $(0,1,1)$
which corresponds to the singular point $(1,0)$ of $G(v,t)=0$.

\begin{figure}[!h]
 \centering
 \includegraphics[width=0.2\textwidth,height=1.2in]{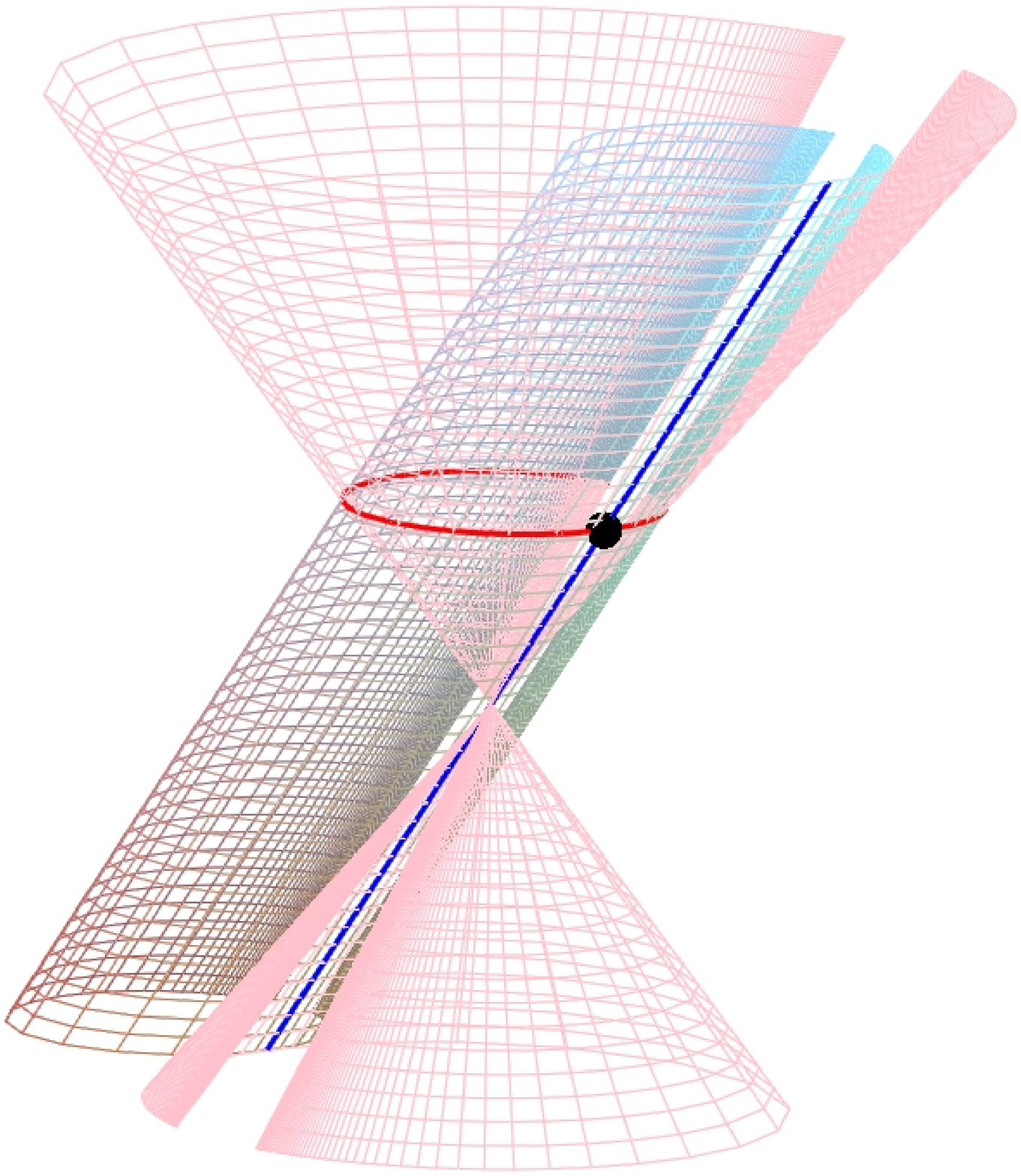}\quad
 \includegraphics[width=0.2\textwidth,height=1.2in]{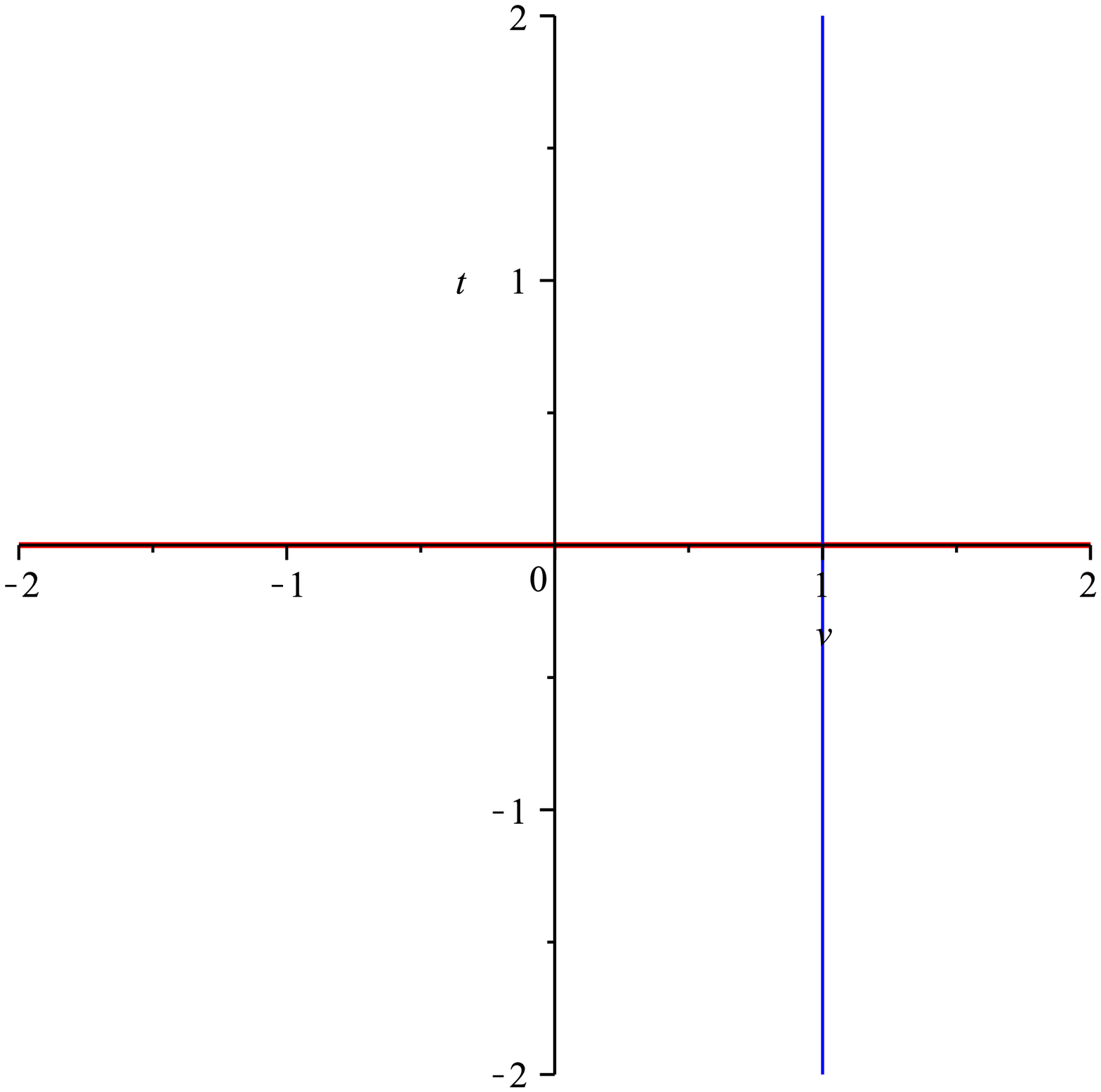}
 \caption{A Cone and an Elliptic
Cylinder / Topology  of the plane curve} \label{figofce}
\end{figure}

This example was involved in a more careful discussion in
\cite{Heo99} and the results are divided to four parts. Comparing
with~\cite{Mario06a}, we add the intersection point $(1,0)$ in the
space $\G^I$. Otherwise, the numerical intersection may consist of
two components separately: a numerical circle no passing through
$(0,1,1)$ and a numerical line passing through $(0,1,1)$.
\end{example}

\begin{example}{\label{ex6}}
As mentioned in Section~\ref{section2}, we construct a tube surface
$$\begin{array}{l}\S_1=\left({s}^{3}+s,{\frac {{s}^{2}+{s}^{2}{u}^{2}+u}{1+{u}^{2}}},\,{\frac {
1-{u}^{2}+2\,s+2\,s{u}^{2}}{2(1+{u}^{2})}} \right),\end{array}$$ formed by a
circle following the space curve $(s^3+s,s^2,s)$. And the surface $\S_2$ is a
whirled surface without much restriction.
$$\begin{array}{l}
\S_2=\left(-\,{\frac { 2( -1+{t}^{2} ) v}{ (
1+{t}^{2} )
 ( 1+{v}^{2} ) }},{\frac {4tv}{ ( 1+{t}^{2} )
( 1+{v}^{2} ) }},{\frac {v ( 3+{v}^{2} ) }{1+{v
}^{2}}}
 \right).
\end{array}$$
The implicit equation of $\S_1$ is
$$\begin{array}{rl}&F(x,y,z)=\\& -1+256\,y{z}^{4}+16\,y+384\,{x}^{2}{z}^{2}y-768\,xz{y}^{3}-52\,{y}^{2}
\\ & +832\,{z}^{2}{x}^{2}+208\,{z}^{4}+192\,{y}^{4}{z}^
{2}+192\,xzy+64\,xz\\ &+64\,{z}^{6}+512\,{y}^{3}{z}^{2} -768
\,{y}^{2}xz-128\,{y}^{3}{x}^{2}-52\,{z}^{2}\\ &-128\,y{z}^{2}
-384\,{x}^{3}z +256\,{y}^{5}-96\,y{x}^{2}+320\,{y}^{2}{x}^{2}\\ & -128\,{y}^{3}-768\,yx{z}^{3}-16\,{x}^{2}+64\,{x}^{4}+64\,{y}
^{6}\\&+192\,{y}^{2}{z}^{4}-768\,{z}^{3}x +208\,{y}^{4}+416\,{y}^
{2}{z}^{2}=0.\end{array}$$
We omit the intersection equation $G(v,t)=0$ for its long expression. The
topology of the plane curve and the numerical intersection curves (red curves)
are illustrated as the following figures (See Fig.~\ref{general2}).
\begin{figure}[!h]
 \centering
 \includegraphics[width=0.2\textwidth,height=1.2in]{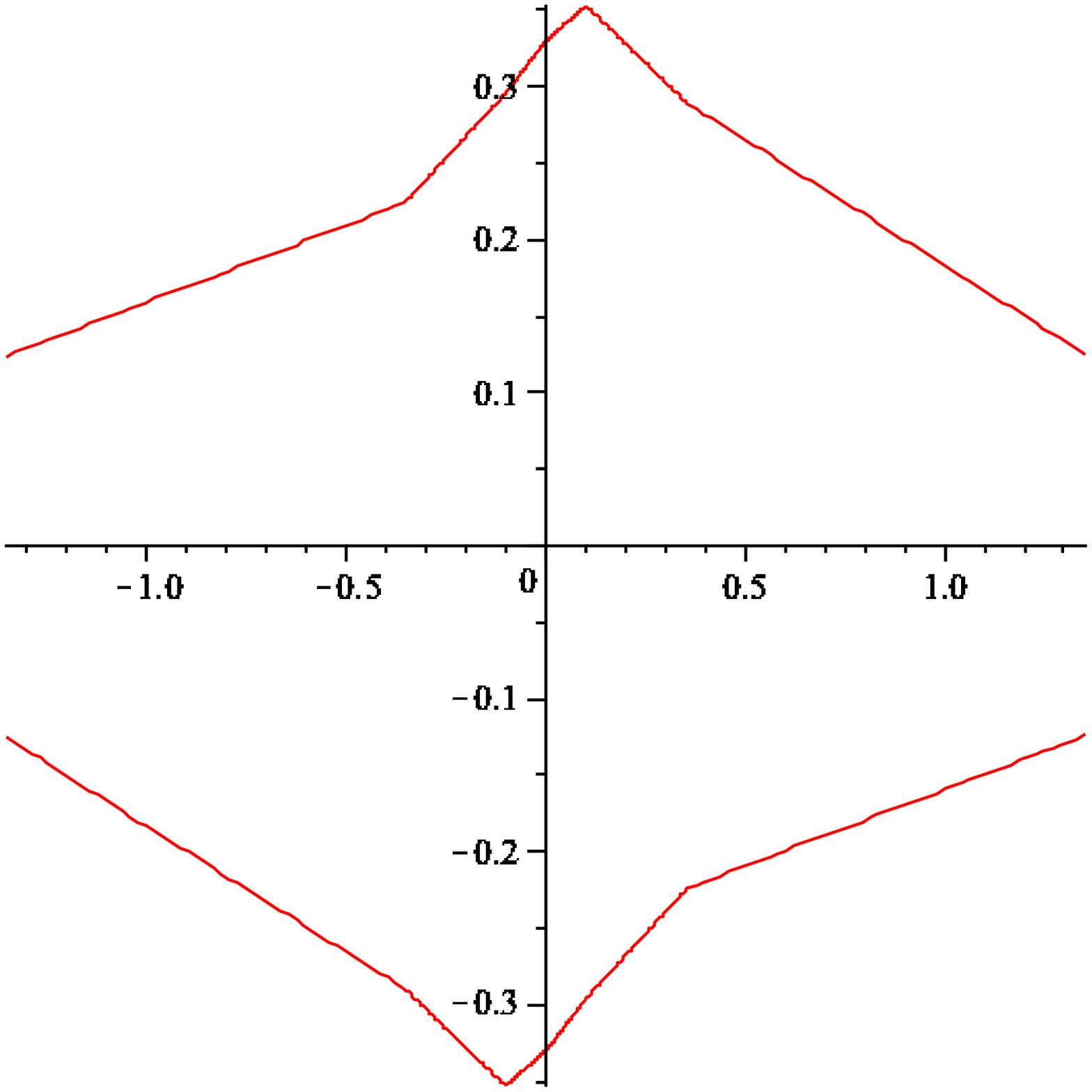}\quad
 \includegraphics[width=0.2\textwidth,height=1.2in]{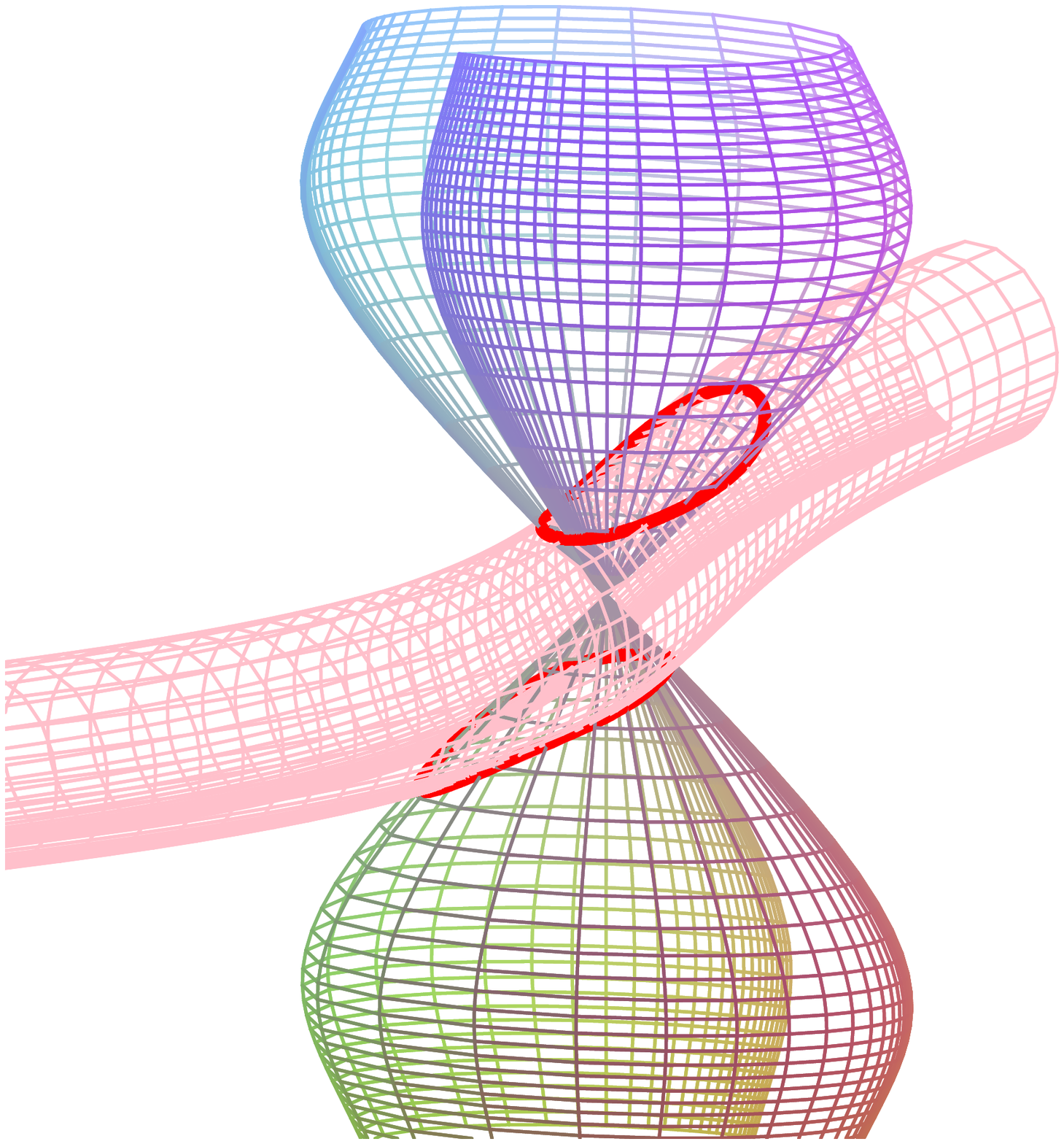}
 \caption{Topology of the plane curve / Numerical intersection curves} \label{general2}
\end{figure}
\end{example}
\begin{example}
The given surfaces are
$$\S_1:\left\{\begin{array}{l}
x=(s u^2+s+1)/2\\
y=su\\
z=u
\end{array}\right.\, \mbox{and} \, \,
\S_2:\left\{\begin{array}{l}
  x={\frac {t \left( 1-{v}^{2} \right) }{1+{v}^{2}}}\\
   y=2\,{\frac {vt}{1+{
v}^{2}}}\\
    z= t
\end{array}\right.$$
$\S_1$ is an elliptic paraboloid whose
implicit equation is $$F(x,y,z)=y^2+z^2-2x+z=0$$ and $\S_2$ is a cone.
Then $G(v,t)=G_1(v,t)G_2(v,t) $ consists of two irreducible factors
as $G_1(v,t)=t$ and $G_2(v,t)=t{v}^{4}+3\,{v}^{4}+6\,t{v}^{2}+t+2\,{v}^{2}t-1$.
It means that the intersection of $\S_1$ and $\S_2$ consists of two components:
$$\mathcal{C}^I_1:\left\{\begin{array}{l}G_1(v,t)=0\\
  x={\frac {t \left( 1-{v}^{2} \right) }{1+{v}^{2}}}\\
   y=2\,{\frac {vt}{1+{
v}^{2}}}\\
    z= t
\end{array}\right.
\quad\mbox{and}\quad
\mathcal{C}^I_2:\left\{\begin{array}{l}G_2(v,t)=0\\
  x={\frac {t \left( 1-{v}^{2} \right) }{1+{v}^{2}}}\\
   y=2\,{\frac {vt}{1+{
v}^{2}}}\\
    z= t
\end{array}\right.
$$
$\mathcal{C}^I_1$ is an isolate point as $(0,0,0)$ and $\mathcal{C}^I_2$ is a quartic space curve with a self-intersected point $(0,0,0)$ with the parameters $\pm(\sqrt{3}/3,0)$.
We can find that the point $(0,0,0)$ plays different roles: 1) an isolate point of $\mathcal{C}_1^I$ corresponding to $(v,0)$ at $\S_2$; 2) a self-intersected point of
$\mathcal{C}_2^I$; 3) the intersect point of $\mathcal{C}_1^I$ and $\mathcal{C}_2^I$; 4) the singular point of $\S_2$.
\begin{figure}[!h]
 \centering
 \includegraphics[width=0.23\textwidth,height=1.2in]{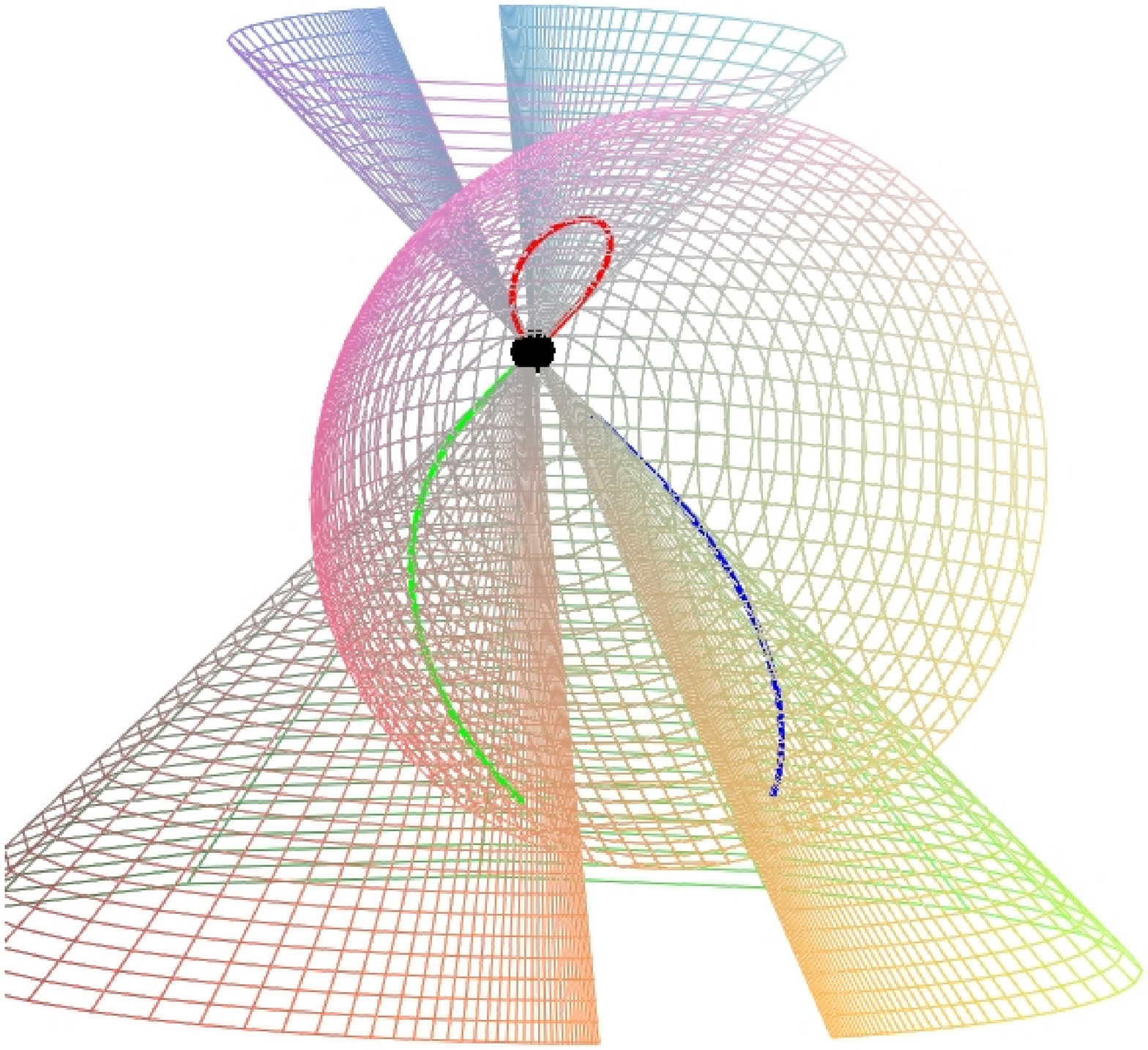}\qquad
 \includegraphics[width=0.20\textwidth,height=1.2in]{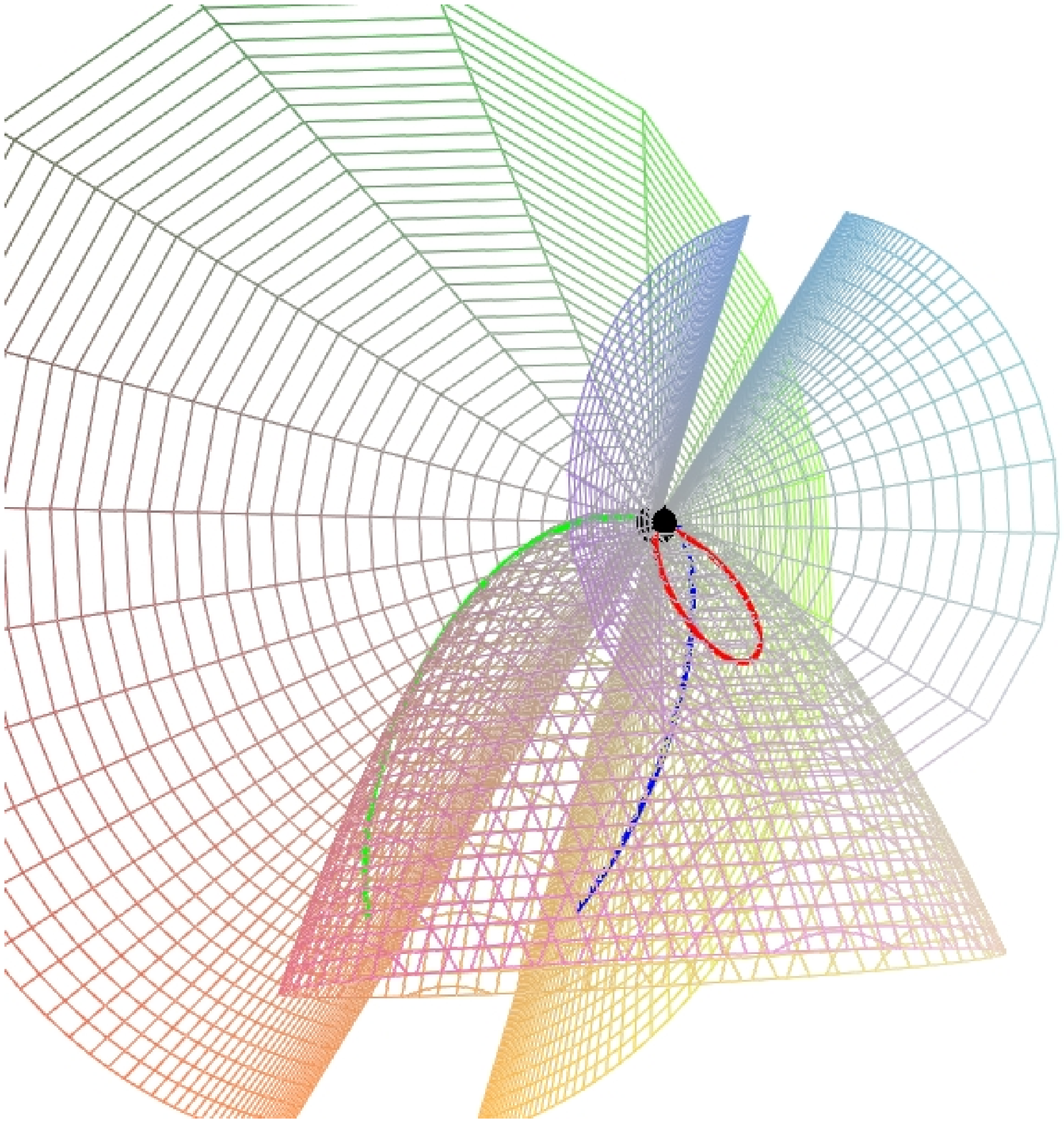}
 \caption{ Numerical intersection curves} \label{con_para}
\end{figure}
\end{example}

\section{Conclusions}
The paper proposes an improved method for approximating the intersection curve of two rational surfaces, one being projectable.
For two given rational surfaces, we
enhance some key steps in the typical process of intersection
analysis. Our method is simpler in implicitization
and adapt to more surfaces, similar enhancement in planar topology
determination. As another important improvement, we refine the topology graph by
adding more singular points as well as their corresponding points of
the intersection curve. Then the space topology graph is homeomorphic
to the intersection curve. And the numerical curve based on the
space topology graph converges to the intersection curve in
subdivision process. Based on the refined topology graph, we can approximate the intersection curve with B-spline curve and we will give the more details in the further paper.

\section*{Acknowledgement} This work is partially supported by National Natural Science Foundation of
China under Grant 10901163, 11001258, a National Key Basic Research Project of China
(2011CB302400) and a President Fund of GUCAS. The authors also wish to
thank the anonymous reviewers for their helpful comments and
suggestions.

\end{document}